\documentclass[cameraready]{Interspeech}
\usepackage{booktabs}
\usepackage{multirow}
\usepackage{amsmath}
\usepackage[table]{xcolor}
\usepackage{subcaption}
\definecolor{cookblue}{HTML}{EEF5FF}
\definecolor{bestgreen}{HTML}{DFF5E1}
\definecolor{gtgray}{HTML}{F2F2F2}

\usepackage{hyperref}

\title{CookVoice: Unified Framework for Style Controllable Multi-Modal Human Voice Generation}

\author[affiliation={1}, orcid=0009-0009-1359-872X]{Haowei}{Lou}
\author[affiliation={1}, orcid=0000-0003-4425-7388]{Hye-Young}{Paik}
\author[affiliation={2}, orcid=0009-0005-7970-846X]{Jia}{Dai}
\author[affiliation={2}, orcid=0000-0002-7260-3105]{Kai}{Li}
\author[affiliation={1}, orcid=0000-0002-4149-839X]{Lina}{Yao}

\address{
    $^1$ UNSW Sydney \\
    $^2$ Dolby Laboratories
}

\email{haowei.lou@unsw.edu.au, h.paik@unsw.edu.au, Jia.Dai@dolby.com, Kai.Li@dolby.com, lina.yao@unsw.edu.au}


\usepackage{comment}

\begin{document}

\maketitle

\begin{abstract} 
Human voice generation has made rapid progress in speech generation, singing voice generation, voice cloning, and voice editing. However, most existing systems are designed for specific tasks and often rely on task-dependent architectures, control signals, or autoregressive decoding, limiting fine-grained controllability and inference efficiency. In this paper, we propose \textbf{CookVoice}, a unified framework for multimodal, multi-style, and multi-task human voice generation. CookVoice decomposes the human voice into three key factors: content, prosody, and style, enabling both speech and singing voice generation within a unified model. To achieve precise and flexible controllability, we design a flexible alignment strategy that maps text, style, and prosody control signals onto the frame-level of spectrogram. This design allows CookVoice to support a wide range of tasks, including text-to-speech, text-to-singing voice, style-controllable generation, voice mimicry, voice conversion, and voice editing. Experimental results show that CookVoice achieves generation quality comparable to existing Text-to-Speech and text-to-singing voice baselines, while providing stronger style and prosody controllability. Moreover, CookVoice achieves comparable performance to large-scale baselines with only \textbf{$43.51$} million parameters and efficient inference using as few as \textbf{4} ODE steps, making it a practical solution for real-world human voice generation applications. Demo page is available at \href{https://haoweilou.github.io/CookVoice/}{https://haoweilou.github.io/CookVoice/}.
\end{abstract}

\section{Introduction}
Recent years have witnessed an emergence of human voice generative AI, with models capable of high-quality text-to-speech~(TTS)~\cite{du2024cosyvoice,deng2025indextts,lou2025parastyletts,chen2025f5}, singing voice generation~\cite{liu2022diffsinger,zhang2024stylesinger,zhang2024tcsinger}, zero-shot voice cloning~\cite{deng2025indextts,chen2025f5,du2024cosyvoice}, and voice editing~\cite{zhang2026bm}. Driven by advances in acoustic representation learning~\cite{kong2020hifi,Lou2026ParaMETATL}, and generative artificial intelligence~\cite{peebles2023scalable,lipman2022flow} modern systems can generate increasingly natural and expressive voices from multi-modal input. However, despite this progress, most existing voice generation systems are still designed around isolated tasks. TTS models focus on generating speech from text~\cite{ren2019fastspeech}, singing voice synthesis models typically require musical scores and note input~\cite{zhang2022m4singer,liu2022diffsinger}, style-controllable TTS models rely on text style prompts or reference speech~\cite{lou2025parastyletts,du2024cosyvoice}, and voice cloning models mainly emphasize speaker similarity~\cite{chen2025f5}. Different forms of human voice generation tasks are commonly treated as separate problems, requiring task-specific architectures, datasets, training objectives, and control signals.

This fragmentation makes human voice generation unnecessarily complex. Theoretically, speech and singing share the same fundamental linguistic basis, yet they differ substantially in temporal structure and prosodic behavior. In speech, duration and prosody are often implicit and context-dependent, while in singing, they are explicitly constrained by musical notes, beats, and melody. Moreover, style-controllable voice generation focuses on cloning the rich paralinguistic styles within reference speech, such as emotion, gender, age, timbre, and speaking or singing style. These styles may come from different modalities, including natural language descriptions, reference speech, reference singing voices, lexical tones, stresses, notes, or F0 contours. Existing methods usually handle only a subset of these controls, making it difficult to flexibly decide \textit{what} to generate, \textit{how} it should sound, and \textit{where} its prosody should come from.

A unified model capable of generating different forms of human voice under a single controllable interface is highly desirable for both academia and the industrial community. Ideally, such a model should support both speech and singing voice generation, accept multimodal style control, and allow explicit control over prosody. This requires a framework that can represent human voice generation through shared factors, including content, prosody, and style, while preserving the distinct characteristics of different tasks. More importantly, the framework should remain simple to train, straightforward to use, and efficient at inference time.

However, achieving such a unified framework remains challenging. Many recent voice generation models rely on autoregressive~(AR) decoding or implicit alignment mechanisms~\cite{du2024cosyvoice,deng2025indextts}, while recent unified voice generation systems further attempt to cover multiple tasks within a single model~\cite{vevo,zhang2026bm}. Although these models can achieve strong generation quality, their controllability is often constrained by the underlying generation mechanism. AR models generate acoustic tokens sequentially and therefore model duration and prosody implicitly through token prediction. This design is flexible at the utterance level, but it makes fine-grained temporal control difficult, since users cannot directly specify how each phoneme, note, or frame should evolve over time. In contrast, many NAR systems~\cite{ren2019fastspeech,lou2025parastyletts,kim2021conditional,chen2025f5} improve temporal stability by imposing explicit or restricted alignments, such as phoneme-level durations~\cite{ren2019fastspeech,kim2021conditional,lou2025parastyletts}, note-to-phoneme mapping~\cite{zhang2022m4singer,zhang2024stylesinger,zhang2024tcsinger}, or sentence-level prosody references~\cite{vevo,zhang2026bm}. While these designs improve controllability over AR decoding, they often rely on predefined assumptions between content and prosody, which limits their flexibility when multiple control signals are jointly involved.

These limitations become more severe for tasks requiring precise temporal control, such as singing voice generation and prosody mimicry. For singing, phonemes must follow score-defined note durations and melodies; for voice mimicry, the generated voice should preserve detailed frame-level prosodic patterns from a reference signal. When content, style, and prosody are combined, implicit AR alignment or restricted NAR alignment makes it difficult to achieve both flexible control and precise temporal consistency. To address this issue, CookVoice adopts a flexible frame-level alignment strategy. Instead of assuming a fixed dependency between content and prosody, CookVoice explicitly expands different control signals to the acoustic frame level and allows them to be combined freely. This design enables more fine-grained and flexible controllability across speech, singing voice generation, and prosody mimicry.

\begin{table}[t]
\centering
\caption{Alignment method and controllability comparison.}
\label{tab:comparison_table}
\begin{tabular}{lcc}
\toprule
\textbf{Model Class} & \textbf{Alignment Method} & \textbf{Controllability} \\
\midrule
AR  & Token prediction & Voice-level \\
NAR & Restricted alignment & Phoneme-level \\
\midrule
CookVoice & Flexible alignment & Frame-level \\
\bottomrule
\end{tabular}
\end{table}

In this work, we propose \textbf{CookVoice}, a unified non-autoregressive~(NAR) framework for multimodal, multi-style, and multi-task human voice generation. CookVoice formulates human voice generation as conditional latent acoustic generation, where the target voice is represented in a compact latent space and generated from three shared factors: content, prosody, and style. Content is represented by phoneme sequences derived from text or lyrics. Prosody can be controlled through discrete signals, such as lexical tones, stresses, and MIDI notes, or continuous frame-level F0 contours extracted from a reference voice. Style can be provided by either text-based descriptions or reference voice prompts, enabling both natural language style control and reference-based style transfer.

Instead of relying on implicit AR token-prediction, CookVoice adopts temporally aligned NAR generation pipeline which explicitly align content, prosody, and style tokens to acoustic tokens at frame level. For speech, phoneme durations are learned or predicted from alignment information. For singing voice, durations are deterministically derived from musical score timing. This design allows speech and singing voices to be modeled under the same temporal framework, while preserving their different duration and prosody structures. The temporally aligned representations are then integrated through a multimodal adaptive fusion module and passed to a flow-matching Diffusion Transformer for high-quality latent acoustic generation.

CookVoice supports a broad range of human voice generation tasks within a single framework, including TTS, text-style-controllable TTS, voice-style-controllable TTS, singing voice generation, style-controllable singing voice generation, prosody mimicry and many others. These tasks are handled by activating different combinations of content, prosody, and style conditions, rather than training separate task-specific models. As a result, CookVoice is a simple, interpretable, and highly controllable framework for the generation of human voice in both speech and singing domains. The main contributions are summarized as follows:
\begin{enumerate}
\item We propose \textbf{CookVoice}, a unified framework for multi-task human voice generation. By decomposing human voice into style, content, and prosody, CookVoice successfully unifying multiple voice generation task within a unified model.

\item We design a flexible alignment mechanism for fine-grained controllable human voice generation. By aligning different control signals at frame level, CookVoice can flexibly support multimodal style and prosody control, enabling more precise control over paralinguistic style, melody, duration, and prosodic expression.

\item Our experiment demonstrate CookVoice achieves higher style and prosody controllability than existing baselines, including models trained with larger-scale data or larger model sizes. Despite using a lightweight architecture and a much smaller training data, CookVoice achieves superior style similarity and F0 controllability while maintaining efficient inference and competitive generation quality.
\end{enumerate}

\section{Preliminary}
\subsection{Problem Formulation}
Human Voice Generation~(HVG) is to generate waveform from different modality of control signal.  In this study, we define the human voice as a combination of human speech and the singing voice. Control signals include \textbf{text}, \textbf{lyric}, \textbf{note}, and \textbf{reference voice}. Full details covered tasks are presented in Appendix~\ref{app:definition}. For readers unfamiliar with linguistic and musical concepts, we provide terminology definitions in Appendix~\ref{app:terminology}.

Formally, we formulate HVG tasks as follow. Let $Y \in \mathbb{R}^{N \times T}$ denotes the latent embedding extracted from a spectrogram via HiFi-GAN~\cite{kong2020hifi}, with $N$ and $T$ denoting the latent dimension and the number of frames. To systematically model the generation process, we assume the human voice can be disentangled into five independent key components:

\noindent \textbf{Style} ($S \in \mathbb{R}^{D}$): Encodes a global paralinguistic styles embedding of voice, indicating how the voice is expressed (e.g., gender, age, and emotion). This style embedding can be conditioned on different sources: either a text description $S_\mathcal{T}$ (e.g., \textit{"A female speaking happily"}), or a reference voice $S_\mathcal{V}$ to clone the speaking style of target voice $\mathcal{V}$;

\noindent \textbf{Content} ($X \in \mathbb{R}^{L_1}$): Represents the linguistic content to be voiced. Content in Text-to-Speech (TTS) can be text, while in Text-To-Singing Voice (TTSV), it refers to the lyrics. In this study, $X$ specifically denotes the sequence of phonemes;

\noindent \textbf{Prosody} ($\mathcal{P}$): contains all forms of prosody variations that shape the expressiveness and melody of the voice. We categorize prosody signals into discrete and continuous signals. The discrete signal ($\mathcal{P}_{disc}$) contains lexical prosody tokens $\mathcal{P}_{lex} \in \mathbb{R}^{L_1}$ (e.g., tones in Mandarin Chinese or word stress in English) for speech, and musical notes $\mathcal{P}_{note} \in \mathbb{R}^{L_2}$ for singing voices. While musical notes govern prosody variation similar to lexical tones, their sequence length ($L_2$) can differ from the phoneme length ($L_1$), as a single musical note may span multiple phonemes, or a single phoneme may stretch across multiple notes. Alternatively, the continuous prosody ($\mathcal{P}_{cont}$) represents the frame-level fundamental frequency ($F_0 \in \mathbb{R}^{T}$) contour extracted from a voice. It aligns temporally with the latent embedding, sharing the same frame length $T$.

Our objective is to train a generative model $\mathcal{G}(\cdot)$ that can generate latent embedding $\hat{Y}$ conditioned on different source of control signal~($\hat{Y} = \mathcal{G}(S, X, \mathcal{P})$). Depending on the requirements of different tasks, $S$ and $\mathcal{P}$ can be selectively conditioned on different source. Detailed task formulations are provided in Table~\ref{tab:task_definition}.

\subsection{Latent Acoustic Encoding}\label{sec:latent_encoding}
We employ an Autoencoder (AE) with similar architecture with the HiFi-GAN~\cite{kong2020hifi} to efficiently encode audio signals into latent embedding. The encoder takes a linear spectrogram $\mathrm{spec} \in \mathbb{R}^{513 \times T}$ as input and compresses it into a continuous latent embedding $Y \in \mathbb{R}^{N \times T}$, where $N$ denotes the latent channel dimension and $T$ represents the temporal frame length. 
The decoder is constructed using a series of transposed 1D convolutional layers, which progressively upsample the latent embedding $Y$ to reconstruct the original audio waveform. To ensure high-fidelity audio generation, the AE is jointly optimized using an adversarial loss from a discriminator alongside a reconstruction loss. 
Once the AE is fully trained, its weights are frozen. Latent embedding $Y$ serves as the ground-truth target for CookVoice during voice generation.

\section{CookVoice}
\begin{figure*}
    \centering
    \includegraphics[width=0.9\linewidth]{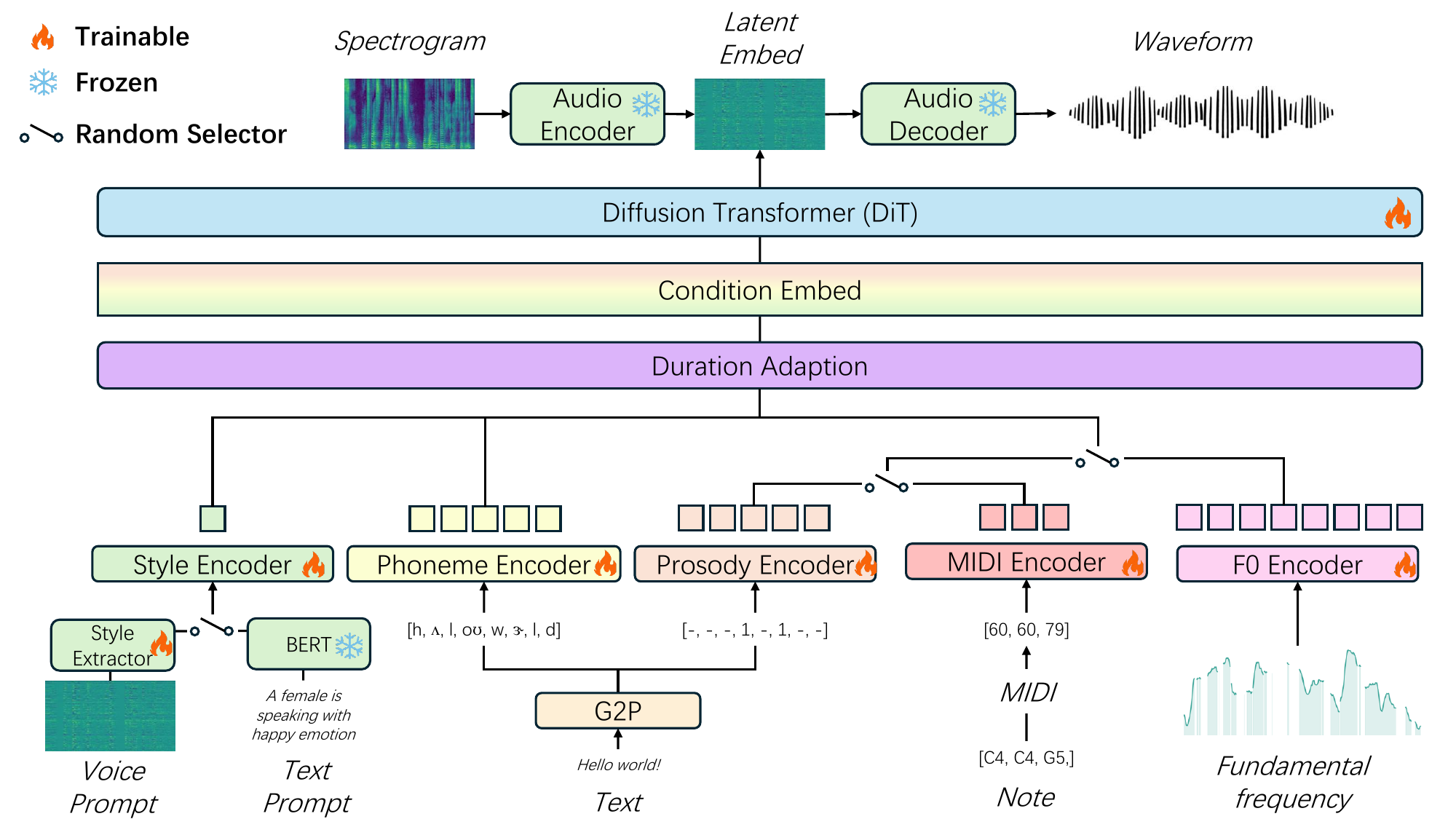}
    \caption{Architecture of the CookVoice}
    \label{fig:cakevoice_overview}
\end{figure*}

The overall architecture of CookVoice is presented in  Figure~\ref{fig:cakevoice_overview}. This section will present the architecture of CookVoice in detail. 
\subsection{Style Encoder}
Style Encoder is designed to generate a style embedding $S$, which dictates the stylistic characteristics of the generated voice. In CookVoice, style conditioning can be derived from two modalities.

The first modality is text, coming from the text-based style prompt $S_\mathcal{T}$. In this scenario, the style is controlled by natural language descriptions (e.g., \textit{a female is speaking with a happy emotion}). Given a text-based style description $\mathcal{T}$, we first extract its semantic embedding using a pre-trained, frozen, MPNet \cite{song2020mpnet} sentence encoder. The resulting embedding is then projected through a trainable linear layer to obtain the text style embedding $S_\mathcal{T} \in \mathbb{R}^{D}$.

The second modality is a reference voice, $\mathcal{V}$. This addresses scenarios requiring zero-shot voice cloning or timbre transfer, where the generated voice must mimic a reference voice. We encode the acoustic signal into a latent space using the encoding method described in Section \ref{sec:latent_encoding}. The resulting latent embedding $\mathcal{V} \in \mathbb{R}^{N \times T}$ is then processed by a trainable Transformer~\cite{vaswani2017attention} encoder with an attentive pooling layer to obtain the voice style embedding $S_\mathcal{V} \in \mathbb{R}^{D}$.

During model training, the overall style condition $S$ is uniformly sampled from either the text modality $S_\mathcal{T}$ or the voice modality $S_\mathcal{V}$ with a 50\% probability for each item within the batch. To align with the temporal resolution of the latent embedding $\mathcal{V} \in \mathbb{R}^{N \times T}$, the style embedding $S$ is expanded by repeating it $T$ times along the temporal axis, producing the sequence-length style embedding $S_e \in \mathbb{R}^{D \times T}$.

\subsection{Content Encoder}
We employ a LanStyleTTS multilingual Grapheme-to-phenome (G2P) module \cite{lou2025generalized} to convert text into phonemes $X$ and prosody tokens $P$, where $X, P \in \mathbb{R}^{L_1}$. The phoneme sequence $X$ captures the linguistic content, while the prosody sequence $P$ explicitly models phoneme-level prosodic variations (e.g., lexical tones in Mandarin or phoneme stress in English) across different linguistic contexts. In CookVoice, $X$ strictly denotes the linguistic content, while $P$ serves as a discrete prosody-conditioning signal (also detailed in Section \ref{sec:discrete_sig}).

Phonemes are first mapped into a sequence of embeddings $X \in \mathbb{R}^{D \times L_1}$ and then processed through Feed-Forward Transformer (FFT) blocks \cite{ren2019fastspeech}. A critical factor for natural and expressive voice generation is the duration of the phoneme $\mathcal{D}_X \in \mathbb{Z}^{L_1}$, which exhibits vastly different behaviors in the speech versus singing voice. In conventional TTS tasks, the absence of explicit duration constraints necessitates the use of a duration predictor \cite{ren2019fastspeech,lou2025parastyletts} or AR architectures \cite{du2024cosyvoice,wang2017tacotron} for implicit inference. Conversely, in TTSV tasks, note durations are explicitly dictated by musical scores. To unify both tasks within CookVoice, we adopt a flexible duration alignment and expansion strategy.

During the training stage, we handle speech and singing voice differently. For speech, we use pre-trained ParaStyleTTS \cite{lou2025parastyletts} to find optimal alignments between the speech audio and the phoneme sequence, yielding ground-truth durations $\mathcal{D}_X$. For singing voice, where score durations are denoted in relative beats rather than absolute time (e.g., seconds), we compute the relative temporal proportion of each phoneme based on its assigned beats. This relative proportion is then multiplied by the total number of acoustic frames $T$ to ascertain the absolute frame-level duration. Both duration sequences satisfy the constraints $\mathcal{D}_X \in \mathbb{Z}^{L_1}$ and $\sum_{i=1}^{L_1} \mathcal{D}_X^i = T$. Each phoneme embedding $X_i$ is then repeated $\mathcal{D}_X^i$ times to produce the duration-expanded content embedding $X_e \in \mathbb{R}^{D \times T}$, which perfectly matches the temporal dimension of the latent acoustic embedding.

During the inference stage, we use a trained duration predictor from  ParaStyleTTS~\cite{lou2025parastyletts} to predict phoneme durations for bilingual speech. For singing voice generation, the duration is deterministically computed from the provided musical score using the relative proportion methodology applied during training.

\subsection{Prosody Encoder}
Prosody affects how the content is expressed differently for speech and singing. In CookVoice, we categorize control signals of prosody, denoted as $\mathcal{P}$, into discrete and continuous signals. The expanded prosody token is denoted as $\mathcal{P}_e \in \mathbb{R}^{D \times T}$.

\subsubsection{Discrete Signals}~\label{sec:discrete_sig}
Discrete prosody signals come from either the G2P module (for speech) or musical scores (for singing). 
For speech, these consist of the aforementioned lexical prosody tokens $P_{lex}$ (tones and stresses). Since the sequence length of $P_{lex}$ matches the phonemes ($L_1$), each token has the same duration $\mathcal{D}^i$ of its corresponding phoneme. Consequently, the mapped token embeddings are repeated $\mathcal{D}^i$ times, giving the expanded discrete speech prosody embedding $P_{disc, e} \in \mathbb{R}^{D \times T}$.

For singing voice, the discrete signals are musical note tokens $P_{note} \in \mathbb{R}^{L_2}$. Similarly to the duration expansion in the Phoneme Encoder, we compute the exact frame duration for each note based on the score's beat allocation and the total frames $T$. The note embeddings are then expanded accordingly, serving as the singing counterpart for $P_{disc, e}$.

\subsubsection{Continuous Signals}
Continuous prosody signals are extracted directly from the reference voice waveform to explicitly represent fundamental frequency variations over time. We extract a continuous $F_0$ contour $\in \mathbb{R}^T$, which is passed through a linear projection layer to generate the continuous prosody embedding $P_{cont} \in \mathbb{R}^{D \times T}$.

During model training, the prosody embedding $\mathcal{P}_e \in \mathbb{R}^{D \times T}$ is constructed by randomly selecting the discrete representation ($P_{disc, e}$) or the continuous representation ($P_{cont}$) with equal probability. This strategy ensures that the model learns robust prosody control from both discrete tokens and continuous signals.

\subsection{Flow-Matching DiT}
Given the style $S_e$, content $X_e$, and prosody $\mathcal{P}_e$ satisfy $S_e,X_e,\mathcal{P}_e \in \mathbb{R}^{D \times T}$, we concatenate them along the feature dimension to construct a conditioning embedding $C \in \mathbb{R}^{3D \times T}$:
$$C = [S_e ; X_e ; \mathcal{P}_e]$$

In the final stage of CookVoice, our objective is to translate the condition embedding $C \in \mathbb{R}^{3D \times T}$ into high-fidelity latent embedding. To achieve this, we use a Diffusion Transformer (DiT) \cite{peebles2023scalable} as the core acoustic model, where $C$ serves as the essential cross-attention conditioning signal to guide the voice generation process.

We frame the generative process within the Optimal Transport Flow-Matching~\cite{lipman2022flow} (OT-FM) paradigm. Assuming a standard Gaussian prior $Y_0 \sim \mathcal{N}(0, I)$ and a target voice latent embedding $Y_1 \in \mathbb{R}^{N \times T}$, our DiT network, parameterized as $v_\theta(Y_t, t, C)$, is trained to regress the target vector field $Y_1 - Y_0$ that drives the linear probability path between the prior and the empirical data distribution. The model is optimized using the standard Flow-Matching Mean Squared Error (MSE) objective.
\begin{equation}
\mathcal{L}_{FM} = \mathbb{E}_{t, Y_0, Y_1} \left[ \left\lVert v_\theta(Y_t, t, C) - (Y_1 - Y_0) \right\rVert^2 \right]
\end{equation}

During inference, we sample the noise $Y_0$ and simulate the empirical probability flow Ordinary Differential Equation (ODE), given by $dY_t = v_\theta(Y_t, t, C) dt$. To solve this ODE and map the unstructured noise to the final acoustic representation $Y_1$, we employ a standard first-order Euler numerical solver. Starting from $t=0$, the trajectory is discretely integrated with a fixed step size $\Delta t$:
\begin{equation}
    Y_{t+\Delta t} = Y_t + \Delta t \cdot v_\theta(Y_t, t, C)
\end{equation}
To transform noise $Y_0$ into latent embedding condition on $C$.

\subsection{Multi-Task Training}
CookVoice supports multiple generation tasks through a simple condition-switching training strategy. Instead of designing task-specific objectives or separate task heads, we randomly replace the conditioning embeddings during training. For each sample in the training batch, the style condition is randomly sampled from either the text-based style embedding $S_{\mathcal{T}}$ or the voice-based style embedding $S_{\mathcal{V}}$. Similarly, the prosody condition is randomly sampled from either the discrete prosody embedding $\mathcal{P}_{disc,e}$ or the continuous $F_0$-based prosody embedding $\mathcal{P}_{cont}$:
\begin{equation}
    S =
    \begin{cases}
    S_{\mathcal{T}}, & p = 0.5, \\
    S_{\mathcal{V}}, & p = 0.5,
    \end{cases}
    \quad
    \mathcal{P}_e =
    \begin{cases}
    \mathcal{P}_{disc,e}, & p = 0.5, \\
    \mathcal{P}_{cont}, & p = 0.5.
    \end{cases}
\end{equation}

This randomization is applied at the sample level within the batch, so different samples may use different style and prosody sources in the same training step. As a result, CookVoice learns to generate voices under different combinations of control signals, such as text style with discrete prosody, voice style with discrete prosody, text style with continuous $F_0$, and voice style with continuous $F_0$. This strategy allows a single model to support multiple speech and singing voice generation tasks without changing the model architecture or training objective.

\section{Experiments}
\subsection{Dataset}
We combine multiple open-sourced bilingual (english and chinese ) speech and singing voice dataset to conduct the experiment including \textit{Baker}~\cite{BakerDataset2020}, \textit{LJSpeech}~\cite{ljspeech17}, \textit{ESD}~\cite{zhou2021seen}, \textit{CREMA-D}~\cite{cao2014crema}, \textit{CommonPhone}~\cite{klumpp2022common}, \textit{Genshin Voice} dataset~\cite{simon3000_genshin_2025}, GTSinger~\cite{zhang2024gtsinger}. The combined dataset contains approximately \textit{123k} voice samples, \textit{168} hours of voices from \textit{6,361} speakers. 

\subsection{Preprocessing}
First, we extract the raw $F_0$ contour (in Hz) from each voice sample. To accommodate the wide and varying pitch ranges inherent to both speech and singing voices, we project the $F_0$ values onto a logarithmic scale and normalize them using a lower of 50 Hz.

\begin{equation}
\tilde{f} = \frac{\log \mathrm{F_0} - \log f_{\min}}{\log f_{\max} - \log f_{\min}}, \quad f_{\min}=50.
\end{equation}
While $\tilde{f}_t$ provides a standardized absolute pitch, relying on it directly can introduce entanglement. Paralinguistic styles, such as age and gender, can effectively dictate a speaker's absolute pitch range and are already captured by the style embedding. Using absolute $F_0$ as an explicit condition would therefore cause interference between prosody and style.
To decouple the melody from the speaker's timbre, we convert the absolute $F_0$ into a relative pitch contour by removing its voice-level mean. Specifically: 
\begin{equation}
\hat{f} =
\begin{cases}
\tilde{f}_t - \mathrm{mean}(\tilde{f}_t) , & t \in \mathcal{U} \\
-2, & t \notin \mathcal{U},
\end{cases}
\end{equation}
$\mathcal{U}$ denotes voiced frames. For unvoiced frames, we set a constant value of $-2$. This design ensures that the explicit $F_0$ input exclusively controls relative intonation and melody, effectively avoiding conflicts with the style.

\subsection{Implementation Details}
The CookVoice model is trained on \textbf{110k} samples, with a singing-to-speech ratio of 1:9. All evaluation samples are not seen by the model during training stage. For each sample, the temporally aligned style, content, and prosody embeddings are concatenated along the feature dimension to form the conditioning embedding. This conditioning embedding is then used as the condition for DiT-S backbone~\cite{peebles2023scalable}, which learns to generate the target latent acoustic embedding. The model is optimized with the Flow-Matching objective~\cite{lipman2022flow}. We train CookVoice on a single NVIDIA RTX 5090 GPU with 32 batch size for 800K steps.

\subsection{Evaluation Metrics}
To comprehensively evaluate the performance of CookVoice, we use both subjective and objective evaluation metrics to assess generated voice quality, intelligibility, style adherence, and prosody accuracy.

\textbf{Subjective Evaluation:} 
We evaluate CookVoice from four perspectives: perceptual quality, content intelligibility, style controllability, and prosody controllability. For subjective evaluation, we conduct Mean Opinion Score~(\textbf{MOS}) listening tests to measure the naturalness and overall perceptual quality of generated speech and singing voices. For singing voice generation, we additionally use Melody Comparative MOS~(\textbf{M-CMOS}) to evaluate melody consistency.

\textbf{Objective Evaluation:} 
We first assess content intelligibility using Word Error Rate~(\textbf{WER}), Phoneme Error Rate~(\textbf{PhoER}), and Prosody Error Rate~(\textbf{ProER}). The generated audio is transcribed by an ASR system, Whisper~\cite{radford2023robust}, and compared with the ground-truth content at different levels. WER measures word-level errors, PhoER measures phoneme-level errors, and ProER measures errors in discrete prosody tokens such as tones, stresses, or note-related prosodic labels.
To evaluate style controllability, we compute Style Similarity~(\textbf{S-SIM}) between the generated voice and the target style reference. Specifically, we extract style embeddings from both audio samples using a pre-trained style encoder~\cite{lou2026autosift} and calculate their cosine similarity, where a higher value indicates better style preservation. To evaluate prosody controllability, we use F0 Root Mean Square Error~(\textbf{F0-RMSE}) and F0 Pearson Correlation~(\textbf{F0-CORR}). F0-RMSE measures the absolute pitch deviation between generated and target F0 contours, while F0-CORR measures whether the generated pitch follows the target pitch trend. Detailed metric definitions are provided in the Appendix~\ref{app:metric_details}.

\begin{table*}[htbp]
\centering
\caption{Controllability evaluation. Include style controllability,  Style-Similarity \textbf{S-SIM~$\uparrow$} and Prosody Controllability \textbf{F0-RMSE}~$\downarrow$ and \textbf{F0-CORR}~$\uparrow$. Values in parentheses indicate the relative performance improvements compared to the best-performing baselines. }
\label{tab:control_table}
\begin{tabular}{ll lc cc}
\toprule
\textbf{Task} & \textbf{Model} & \textbf{Style + Prosody} & \textbf{SSIM ($\uparrow$) \%}  & \textbf{F0-RMSE ($\downarrow$)} & \textbf{F0-CORR ($\uparrow$)} \\
\midrule
\multirow{12}{*}{TTS}
& CosyVoice~\cite{du2024cosyvoice} & Text + DISC &  30.53 $\pm$ 10.98 & 138.18 $\pm$ 43.17 & 0.1065 $\pm$ 0.1709 \\
& CosyVoice~\cite{du2024cosyvoice} & Voice + DISC & 46.67 $\pm$ 17.10 & 126.35 $\pm$ 45.23 & 0.1902 $\pm$ 0.2423 \\ 
& F5-TTS~\cite{chen2025f5} & Voice + DISC & 66.29 $\pm$ 13.73 & 136.56 $\pm$ 46.28 & 0.1505 $\pm$ 0.1703 \\ 
& ParaStyleTTS~\cite{lou2025parastyletts} & Text + DISC & 42.96 $\pm$ 17.73 & 126.56 $\pm$ 37.93 & 0.1777 $\pm$ 0.2206 \\
& IndexTTS~\cite{deng2025indextts} & Voice + DISC & 71.19 $\pm$ 13.77 & 123.65 $\pm$ 43.99 & 0.2548 $\pm$ 0.2560 \\
& Vevo2~\cite{zhang2026bm} & Voice + DISC &  75.11 $\pm$ 8.20 & 136.00 $\pm$ 45.09 & 0.1242 $\pm$ 0.1806 \\
\cmidrule(lr){2-6}
& CookVoice &   \\
\cmidrule(lr){2-6}
& \quad $S_{\mathcal{T}} + \mathcal{P}_{disc}$ & Text + DISC & 60.78 $\pm$ 14.86$_{(+41.48\%)}$ & 119.56 $\pm$ 50.73$_{(\downarrow 5.53\%)}$ & 0.3929 $\pm$ 0.2395$_{(+121.10\%)}$\\
& \quad $S_{\mathcal{T}} + \mathcal{P}_{cont}$ & Text + CONT & 73.09 $\pm$ 16.38 & 88.85 $\pm$ 38.56 & 0.6416 $\pm$ 0.2011 \\ 
& \quad $S_{\mathcal{V}} + \mathcal{P}_{disc}$ & Voice + DISC & 87.47 $\pm$ 3.77$_{(+16.46\%)}$ & 99.05 $\pm$ 43.77$_{(\downarrow 19.89\%)}$ & 0.5330 $\pm$ 0.2338$_{(+109.18\%)}$ \\
& \quad $S_{\mathcal{V}} + \mathcal{P}_{cont}$ & Voice + CONT & 91.65 $\pm$ 2.98 & 74.98 $\pm$ 24.53 & 0.7102 $\pm$ 0.1714 \\
\midrule
\multirow{10}{*}{TTSV}
& DiffSinger~\cite{liu2022diffsinger} & Text + DISC & 75.81 $\pm$ 5.81 & 100.48 $\pm$ 26.23 & 0.4580 $\pm$ 0.1857 \\
& StyleSinger~\cite{zhang2024stylesinger} & Voice + DISC & 82.00 $\pm$ 5.40 & 101.82 $\pm$ 25.79 & 0.4856 $\pm$ 0.1518 \\
& TCSinger~\cite{zhang2024tcsinger} & Voice + DISC & 87.53 $\pm$ 5.02 & 92.50 $\pm$ 22.01 & 0.5643 $\pm$ 0.1749 \\
& Vevo1.5~\cite{vevo} & Voice + CONT & 86.55 $\pm$ 7.18 & 79.56 $\pm$ 28.69 & 0.7125 $\pm$ 0.1615 \\
& Vevo2~\cite{zhang2026bm}  & Voice + CONT & 88.09 $\pm$ 3.56 & 90.09 $\pm$ 26.94 & 0.6242 $\pm$ 0.1479  \\
\cmidrule(lr){2-6}
& CookVoice &   \\
\cmidrule(lr){2-6}
& \quad $S_{\mathcal{T}} + \mathcal{P}_{disc}$ & Text + DISC & 85.75 $\pm$ 5.41$_{(+13.11\%)}$ & 97.73 $\pm$ 27.37$_{(\downarrow 2.74\%)}$ & 0.5383 $\pm$ 0.1825$_{(+17.53\%)}$ \\
& \quad $S_{\mathcal{T}} + \mathcal{P}_{cont}$ & Text + CONT & 72.51 $\pm$ 18.02 & 68.12 $\pm$ 20.69 & 0.8401 $\pm$ 0.0655 \\ 
& \quad $S_{\mathcal{V}} + \mathcal{P}_{disc}$ & Voice + DISC & 93.72 $\pm$ 2.00$_{(+7.07\%)}$ & 95.01 $\pm$ 28.43$_{(\uparrow 2.71\%)}$ & 0.5704 $\pm$ 0.1685$_{(+1.08\%)}$ \\ 
& \quad $S_{\mathcal{V}} + \mathcal{P}_{cont}$ & Voice + CONT & 95.00 $\pm$ 1.72$_{(+7.84\%)}$ & 56.96 $\pm$ 14.76$_{(\downarrow 28.41\%)}$ & 0.8425 $\pm$ 0.0711$_{(+18.25\%)}$ \\

\bottomrule
\end{tabular}
\end{table*}

\begin{table*}[htbp]
\centering
\caption{Quality evaluation in terms of Mean Opinion Score (\textbf{MOS}), Melody Comparative Mean Opinion Score (\textbf{MC-MOS}) and Word Error Rate~(\textbf{WER}) for Chinese (\textbf{CH}) and English (\textbf{EN}).}
\label{tab:quality_table}
\begin{tabular}{lllcccc}
\toprule
\textbf{Task} & \textbf{Model} & \textbf{Style + Prosody} & \textbf{MOS} $\uparrow$ & \textbf{MC-MOS} $\uparrow$ & \textbf{WER-CH} $\downarrow$ & \textbf{WER-EN} $\downarrow$ \\
\midrule

\multirow{12}{*}{TTS}
& CosyVoice~\cite{du2024cosyvoice}  & Text + DISC  & 4.30 $\pm$ 0.90 & -- &  5.97 $\pm$ 7.53 & 3.14 $\pm$ 6.11  \\
& CosyVoice~\cite{du2024cosyvoice}  & Voice + DISC & 4.05 $\pm$ 1.07 & -- &  11.27 $\pm$ 12.40 & 5.79 $\pm$ 9.85  \\
& F5-TTS~\cite{chen2025f5} & Voice + DISC & 4.35 $\pm$ 0.96 & -- & 5.23 $\pm$ 7.45 & 2.34 $\pm$ 4.67 \\
& ParaStyleTTS~\cite{lou2025parastyletts} & Text + DISC & 4.03 $\pm$ 1.01 & -- & 9.67 $\pm$ 8.25 & 5.21 $\pm$ 6.95 \\
& IndexTTS~\cite{deng2025indextts} & Voice + DISC & 4.42 $\pm$ 0.80  & -- & 4.16 $\pm$ 6.63 & 2.88 $\pm$ 6.04 \\
& Vevo2~\cite{zhang2026bm} & Voice + DISC & 4.30 $\pm$ 0.75 & -- & 6.35 $\pm$ 6.73 & 3.82 $\pm$ 6.47 \\
\cmidrule(lr){2-7}
& \multicolumn{3}{l}{\textit{CookVoice}} \\
& \quad $S_{\mathcal{T}} + \mathcal{P}_{disc}$ & Text + DISC & 3.45 $\pm$ 1.14 & -- &  9.48 $\pm$ 11.88 & 6.07 $\pm$ 12.35 \\
& \quad $S_{\mathcal{T}} + \mathcal{P}_{cont}$ & Text + CONT & 3.80 $\pm$ 1.08 & -- &  9.03 $\pm$ 10.98 & 5.23 $\pm$ 8.76 \\
& \quad $S_{\mathcal{V}} + \mathcal{P}_{disc}$ & Voice + DISC & 3.60 $\pm$ 1.02 & -- &  8.66 $\pm$ 10.14 & 4.41 $\pm$ 7.40 \\
& \quad $S_{\mathcal{V}} + \mathcal{P}_{cont}$ & Voice + CONT & 3.98 $\pm$ 0.91 & -- &  8.13 $\pm$ 10.07 & 4.19 $\pm$ 7.45 \\
\cmidrule(lr){2-7}
\rowcolor{gtgray}
& Ground Truth & & 4.05 $\pm$ 0.92 & -- & 6.33 $\pm$ 9.33 & 3.39 $\pm$ 7.07 \\

\midrule

\multirow{11}{*}{TTSV}
& DiffSinger~\cite{liu2022diffsinger} & Text + DISC & 2.80 $\pm$ 0.81  & -0.80 $\pm$ 1.66 &  17.79 $\pm$ 14.16 & -- \\
& StyleSinger~\cite{zhang2024stylesinger} & Voice + DISC & 3.20 $\pm$ 1.08  & -0.40 $\pm$ 1.61 & 11.31 $\pm$ 7.68 & -- \\
& TCSinger~\cite{zhang2024tcsinger} & Voice + DISC & 2.40 $\pm$ 1.07  & -0.95 $\pm$ 1.41 &  9.88 $\pm$ 7.23 & 29.03 $\pm$ 19.31\\
& Vevo1.5~\cite{vevo} & Voice + CONT  & 3.33 $\pm$ 1.29 & -0.74 $\pm$ 1.66 &  8.85 $\pm$ 11.09 & 23.62 $\pm$ 14.67 \\
& Vevo2~\cite{zhang2026bm} & Voice + CONT & 3.42 $\pm$ 1.30 & 0.11 $\pm$ 1.76 &  4.66 $\pm$ 6.04 & 11.37 $\pm$ 9.80 \\
\cmidrule(lr){2-7}
& \multicolumn{3}{l}{\textit{CookVoice}} \\
& \quad $S_{\mathcal{T}} + \mathcal{P}_{disc}$ & Text + DISC & 2.40 $\pm$ 0.89 & -1.39 $\pm$ 1.24 & 11.71 $\pm$ 10.44 & 19.56 $\pm$ 14.47 \\
& \quad $S_{\mathcal{T}} + \mathcal{P}_{cont}$ & Text + CONT & 3.20 $\pm$ 0.93 & -0.19 $\pm$ 1.44 & 9.79 $\pm$ 9.26 & 21.10 $\pm$ 17.81 \\
& \quad  $S_{\mathcal{V}} + \mathcal{P}_{disc}$ & Voice + DISC & 2.80 $\pm$ 1.00  &  -0.94 $\pm$ 1.41 & 11.45 $\pm$ 9.14 & 22.53 $\pm$ 16.92 \\
& \quad $S_{\mathcal{V}} + \mathcal{P}_{cont}$ & Voice + CONT & 3.40 $\pm$ 0.97 & 0.28 $\pm$ 1.44 & 10.01 $\pm$ 9.70 & 24.37 $\pm$ 17.23\\
\cmidrule(lr){2-7}
\rowcolor{gtgray}
& Ground Truth &  & 3.95 $\pm$ 0.89& -- &  5.01 $\pm$ 4.76 & 14.22 $\pm$ 12.93 \\

\bottomrule
\end{tabular}
\end{table*}

\begin{table*}[t]
\centering
\scriptsize
\setlength{\tabcolsep}{2.8pt}
\renewcommand{\arraystretch}{1.08}
\caption{
Capability--performance--efficiency comparison table with representative voice
generation baselines. The task and control columns summarize the configurations
evaluated in this study. T and V denote text- and voice-based style control,
while D and C denote discrete and continuous prosody control, respectively.
Performance values report the best observed result among the evaluated
configurations of each model. S-SIM and F0-CORR are presented as TTS/TTSV.
}
\label{tab:overall_tradeoff}
\begin{tabular}{lccc cc cc ccc}
\toprule
& \multicolumn{3}{c}{\textbf{Functional Coverage}}
& \multicolumn{2}{c}{\textbf{Perceptual Quality}}
& \multicolumn{2}{c}{\textbf{Controllability}}
& \multicolumn{3}{c}{\textbf{Computational Efficiency}} \\
\cmidrule(lr){2-4}
\cmidrule(lr){5-6}
\cmidrule(lr){7-8}
\cmidrule(lr){9-11}

\textbf{Model}
& \textbf{Task}
& \textbf{Style}
& \textbf{Prosody}
& \shortstack{\textbf{TTS}\\\textbf{MOS}}
& \shortstack{\textbf{TTSV}\\\textbf{MOS / MC-MOS}}
& \shortstack{\textbf{S-SIM (\%)}\\\textbf{TTS / TTSV}}
& \shortstack{\textbf{F0-CORR}\\\textbf{TTS / TTSV}}
& \textbf{\#Para}
& \shortstack{\textbf{CUDA}\\\textbf{Mem.}}
& \textbf{RTF} \\
\midrule

CosyVoice~\cite{du2024cosyvoice}
& TTS & T+V & D
& 4.30 & -- / --
& 46.67 / --
& 0.1902 / --
& 436.00M & 2.05G & 3.74 \\

F5-TTS~\cite{chen2025f5}
& TTS & V & D
& 4.35 & -- / --
& 66.29 / --
& 0.1505 / --
& 300.00M & 2.15G & 6.15 \\

ParaStyleTTS~\cite{lou2025parastyletts}
& TTS & T & D
& 4.03 & -- / --
& 42.96 / --
& 0.1777 / --
& 52.51M & 1.22G & \textbf{0.03} \\

IndexTTS~\cite{deng2025indextts}
& TTS & V & D
& \textbf{4.42} & -- / --
& 71.19 / --
& 0.2548 / --
& 608.00M & 1.53G & 0.36 \\

Vevo2~\cite{zhang2026bm}
& TTS+TTSV & V & D+C
& 4.30 & \textbf{3.42} / 0.11
& 75.11 / 88.09
& 0.1242 / 0.6242
& 872.00M & 6.84G & 14.85 \\

\midrule

DiffSinger~\cite{liu2022diffsinger}
& TTSV & T & D
& -- & 2.80 / $-0.80$
& -- / 75.81
& -- / 0.4580
& \textbf{26.74M} & \textbf{0.60G} & 0.34 \\

StyleSinger~\cite{zhang2024stylesinger}
& TTSV & V & D
& -- & 3.20 / $-0.40$
& -- / 82.00
& -- / 0.4856
& 42.00M & 0.88G & 0.36 \\

TCSinger~\cite{zhang2024tcsinger}
& TTSV & V & D
& -- & 2.40 / $-0.95$
& -- / 87.53
& -- / 0.5643
& 329.50M & 1.97G & 0.33 \\

Vevo1.5~\cite{vevo}
& TTSV & V & C
& -- & 3.33 / $-0.74$
& -- / 86.55
& -- / 0.7125
& 1468.43M & 9.44G & 9.83 \\

Vevo2~\cite{zhang2026bm}
& TTS+TTSV & V & D+C
& 4.30 & \textbf{3.42} / 0.11
& 75.11 / 88.09
& 0.1242 / 0.6242
& 872.00M & 6.84G & 14.85 \\
\midrule

\rowcolor{cookblue}
\textbf{CookVoice}
& \textbf{TTS+TTSV}
& \textbf{T+V}
& \textbf{D+C}
& 3.98
& 3.40 / \textbf{0.28}
& \textbf{91.65 / 95.00}
& \textbf{0.7102 / 0.8425}
& 43.51M
& 1.37G
& 0.04 \\

\bottomrule
\end{tabular}
\end{table*}

\section{Results \& Discussion}\label{sec:result}

In this section, we provide an analysis of experimental results and organize the discussion to answer the following five research questions~(\textbf{RQs}):

\begin{itemize}
\item \textbf{RQ1:} How does CookVoice perform compared with existing TTS and TTSV baselines in term of audio?
\item \textbf{RQ2:} How controllable is CookVoice in terms of style and prosody compared with existing baselines?
\item \textbf{RQ3:} How do different style and prosody control signals affect the generation performance?
\item \textbf{RQ4:} How does the number of ODE solver steps affect the quality-efficiency trade-off?
\item \textbf{RQ5:} How efficient is CookVoice compared with existing TTS and TTSV models?
\end{itemize}


\subsection{RQ1: Performance Comparison}

We compare CookVoice with representative TTS and TTSV baselines in terms of controllability, perceptual quality, and intelligibility using both subjective and objective metrics. As shown in Tables~\ref{tab:control_table} and \ref{tab:quality_table}, CookVoice achieves comparable generation quality while providing stronger style and prosody controllability across both TTS and TTSV tasks.

The main advantage of CookVoice lies in controllability. For TTS, CookVoice consistently improves style similarity and prosody consistency over the compared baselines. Under voice-based style control, CookVoice achieves an S-SIM of over $90\%$ and an F0-CORR of up to $0.7102$, whereas the best TTS baselines (Vevo2) achieve an SSIM of approximately $75\%$ and an F0-CORR of approximately $0.25$. A similar trend is observed for TTSV, CookVoice achieves the highest style similarity and prosody consistency, reaching an S-SIM of $95.00\%$ and an F0-CORR of $0.8425$. These results demonstrate that CookVoice more accurately preserves the style instruction from either voice or text modality and follows the prosody sourced from different types of control signals.

In terms of perceptual quality, CookVoice exhibits different performance patterns for speech and singing voice generation. For TTS, its MOS scores are lower than most TTS baselines, although its best configuration achieves a MOS of $3.98$, close to the ground-truth score of $4.05$. For TTSV, CookVoice achieves a best MOS of $3.40$, outperforming most of the compared singing voice generation systems and remaining comparable to Vevo2, which achieves $3.42$. CookVoice also obtains the highest MC-MOS score, indicating better perceived melody consistency. 

For intelligibility, CookVoice achieves comparable WER results. Its TTS performance lies in the middle range of the evaluated baselines. For TTSV, CookVoice outperforms most baselines on the Chinese song and achieves the second-lowest English WER among the models that support English singing voice generation. One possible reason for the generally higher WERs in TTSV is that Whisper is primarily trained on large-scale speech data. Singing voices exhibit different prosodic patterns, including wider pitch variations and prolonged phonemes, which may increase recognition errors. CookVoice achieves comparable performance with existing baseline who is trained on substantially larger datasets. Its primary advantages lie in its stronger controllability and flexible support for different style and prosody control signals.

\subsection{RQ2: Controllability Analysis}
As shown in Table~\ref{tab:control_table}, existing baselines typically support only a limited combination of style and prosody conditions. In contrast, CookVoice accommodates both text- and voice-based style control together with discrete prosody and continuous $F_0$ control. These control signals are explicitly aligned with the acoustic representation at the frame level, enabling finer-grained guidance throughout the generation process.

Compared with baselines under matched style--prosody settings, CookVoice delivers clear improvements. For TTS with text-based style and discrete prosody control, CookVoice improves S-SIM by $41.48\%$ and F0-CORR by $121.10\%$, while reducing F0-RMSE by $5.53\%$. With voice-based style and discrete prosody control, CookVoice improves S-SIM by $16.46\%$ and F0-CORR by $109.18\%$, while reducing F0-RMSE by $19.89\%$. These results suggest that CookVoice benefits from its explicit frame-level conditioning design. Outperforming existing voice cloning methods that rely primarily on autoregressive token prediction or voice-level style conditioning.

A similar trend is observed for TTSV. Under text-based style and discrete note control, CookVoice improves SSIM by $13.11\%$ and F0-CORR by $17.53\%$, while reducing F0-RMSE by $2.74\%$. Under voice-based style and continuous $F_0$ control, it improves S-SIM by $7.84\%$ and F0-CORR by $18.25\%$, while reducing F0-RMSE by $28.41\%$. 
The higher level of controllability of CookVoice can be attributed to the joint effect of its multimodal control interfaces and fine-grained frame-level conditioning. Together, these designs enable a broader range of controllable generation functions while improving both style preservation and prosody-following consistently across speech and singing voice generation.

\subsection{RQ3: Effects of Different Control Signals}
\begin{figure}
    \centering
    \begin{subfigure}{0.49\linewidth}
        \centering
        \includegraphics[width=\linewidth]{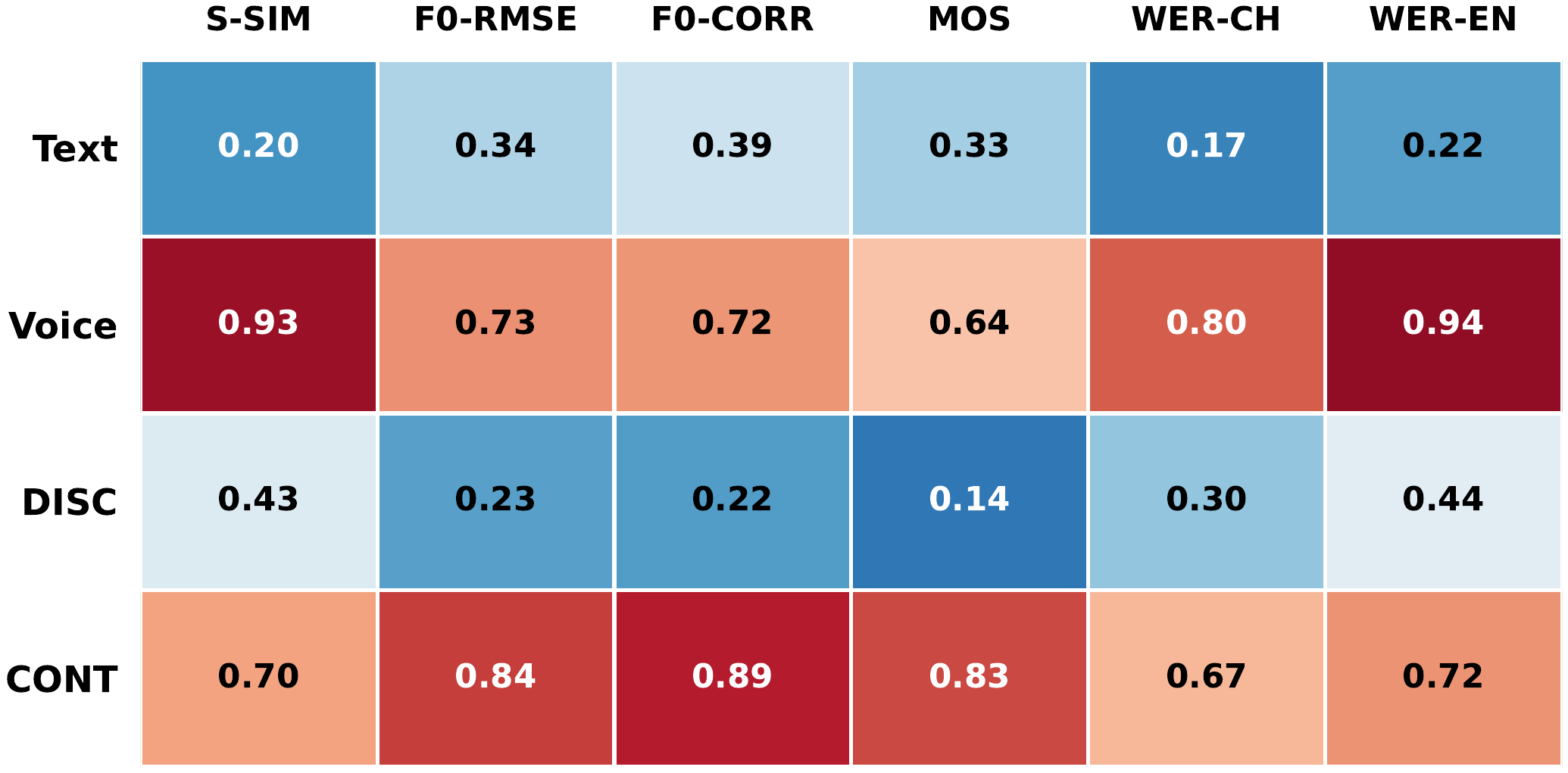}
        \caption{TTS}
        \label{fig:heatmap_tts}
    \end{subfigure}
    \begin{subfigure}{0.49\linewidth}
        \includegraphics[width=\linewidth]{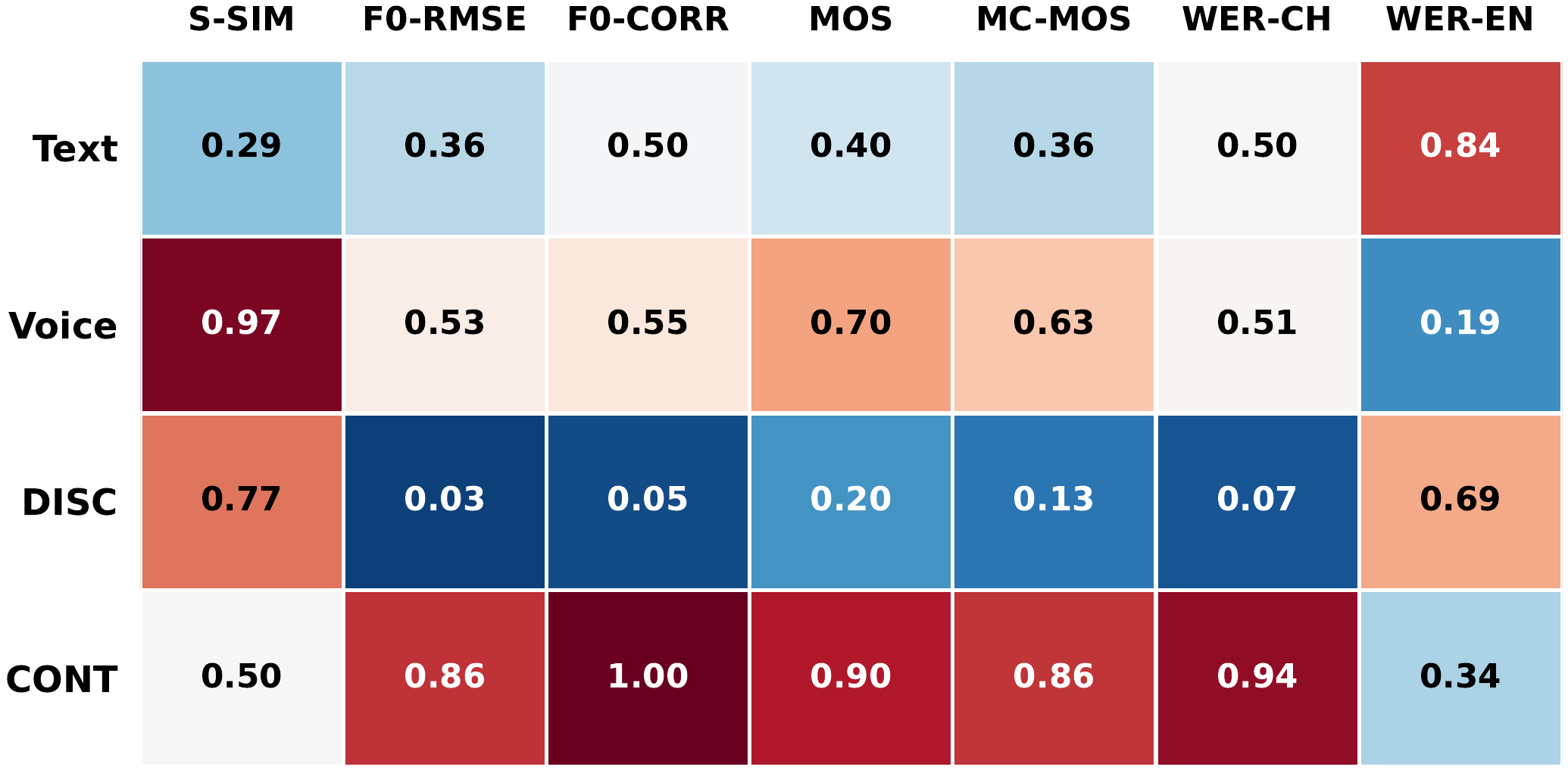}
        \caption{TTSV}
        \label{fig:heatmap_ttsv}
    \end{subfigure}
    \caption{Marginal effects of different style and prosody control signals on CookVoice's performance.  Redder indicate stronger performance, bluer indicate weaker performance.}
    \label{fig:heatmap}
\end{figure}
In this section, we analyze how different control signals affect the performance of CookVoice.
Figure~\ref{fig:heatmap} summarizes the marginal effects of different style and prosody control signals. Each value represents the average normalized performance of one control source across its matched performance, a higher value indicates better performance. The results reveal effects from the style source and the prosody source. 

\noindent \textbf{Effect of the style control source.}
For TTS in Figure~\ref{fig:heatmap_tts}, voice-based style conditioning consistently outperforms text-based conditioning across all evaluated metrics. The largest difference is observed in SSIM, where Voice achieves a normalized score of $0.932$, compared with $0.199$ for Text. Voice conditioning also produces better prosody consistency, perceptual quality, and intelligibility, achieving substantially higher scores for F0-CORR, MOS, and both Chinese and English WER. Voices provide richer acoustic information than natural language style prompts. It allows the model to preserve paralinguistic details such as timbre, emotion and age from the reference voice more accurately. 

As shown in figure~\ref{fig:heatmap_ttsv}. Voice conditioning achieves the highest SSIM and improves both MOS and MC-MOS, confirming its advantage for accurate singing-style transfer. However, its effect on intelligibility is less consistent. Voice and Text have similar Chinese WER performance, while Text achieves a substantially better English WER score. Although voice references improve style preservation and perceptual quality, stronger acoustic style imitation does not necessarily improve lyric intelligibility, particularly for English singing.

\noindent  \textbf{Effect of the prosody control source.}
Continuous prosody signals $F_0$ provides a clear advantage over discrete prosody signals in prosody-related metrics. For TTS, CONT achieves scores of $0.844$ for F0-RMSE and $0.892$ for F0-CORR, compared with $0.230$ and $0.221$ for DISC. It also improves MOS and WER, indicating that detailed frame-level pitch guidance benefits not only prosody following but also overall speech quality and intelligibility.

The advantage of continuous control is even more pronounced for TTSV. CONT substantially outperforms DISC in F0-RMSE, F0-CORR, MOS, MC-MOS, and Chinese WER, confirming that explicit frame-level $F_0$ trajectories are effective for melody following and perceptual singing quality. However, unlike in TTS, DISC achieves higher S-SIM and better English WER performance in TTSV. One possible explanation is that the discrete prosody signal in TTSV consists of musical note, which already provide a structured representation of pitch, duration, and note boundaries that is closely aligned with the lyrics. This flexibility can help preserve the style and maintain clearer lyric articulation. Conversely, continuous signals contain detailed prosody trajectory from the reference voice. Which leads to an observable style-sim degradation under text-based style conditioning, where replacing discrete notes with continuous $F_0$ reduces SSIM from $85.75\%$ to $72.51\%$.

\subsection{RQ4: Effect of ODE Steps}
We further analyze how performance is affected by the number of ODE solver steps, with detailed results provided in Appendix~\ref{sec:analysis}. Increasing the number of ODE steps generally improves style similarity and prosody fidelity, particularly under voice-based style conditioning and continuous prosody control.

Due to the high cost of collecting MOS and MC-MOS ratings, we evaluate these subjective metrics at coarser intervals, increasing the number of ODE steps by a factor of four, while the objective metrics are evaluated at each doubling of the step count. MOS continues to improve up to 16 steps before declining, whereas MC-MOS reaches its highest value at 4 steps and gradually decreases thereafter. Meanwhile, style similarity and $F_0$-based metrics generally stabilize between 4 and 8 steps, and intelligibility metrics such as WER, PhoER, and ProER achieve their best or near-best results at approximately 4 steps and may slightly degrade with further refinement. These results indicate that additional ODE steps can improve acoustic naturalness and prosodic details, but do not necessarily benefit melody consistency or linguistic content preservation.

Based on these observations, we suggest 4--8 ODE steps as the optimal operating range for CookVoice. Within this range, the generated voices maintain stable controllability, good perceptual quality, and intelligibility without incurring additional inference steps.

\subsection{RQ5: Efficiency Analysis}

Table~\ref{tab:overall_tradeoff} provides an integrated comparison of
functional coverage, generation performance, controllability, and
computational efficiency. The results reveal that existing systems typically
optimize only a subset of these dimensions. Large-scale TTS models such as
F5-TTS and IndexTTS achieve higher speech MOS scores than CookVoice, but are
restricted to speech generation and require approximately $6.9\times$ and
$14.0\times$ more parameters, respectively. Similarly, lightweight
task-specific models such as ParaStyleTTS, DiffSinger, and StyleSinger provide
competitive efficiency, but support only speech or singing voice generation
and a limited set of control signals.

Vevo2 is the most directly comparable baseline because it supports both speech
and singing voice generation. Vevo2 achieves higher TTS MOS and a marginally
higher TTSV MOS than CookVoice. However, CookVoice obtains a higher MC-MOS and
substantially stronger style and prosody controllability across both tasks.
More importantly, CookVoice uses only $4.99\%$ of the parameters, $20.03\%$ of
the CUDA memory, and $0.27\%$ of the RTF required by Vevo2.

These comparisons indicate that the main advantage of CookVoice does not lie
in maximizing a single task-specific quality metric. Instead, CookVoice
provides a favorable trade-off among functional coverage, fine-grained
controllability, perceptual quality, and computational efficiency. With a
single $43.51$M-parameter model, it supports both TTS and TTSV, text- and
voice-based style conditioning, and discrete and continuous prosody control,
while maintaining real-time generation with an RTF of $0.04$.

The compact model size and fast inference of CookVoice mainly stem from its
streamlined architecture. By replacing autoregressive token prediction with a
lightweight DiT-S generative backbone, CookVoice limits the model size to
$43.51$M parameters and avoids the sequential next-token decoding required by
AR models. In addition, CookVoice employs flow matching, which requires only a
small number of ODE solver steps, compared with conventional diffusion
sampling processes that typically involve hundreds of iterative denoising
steps~\cite{ho2020denoising}. These design choices reduce both model complexity
and inference cost.

\section{Conclusion}
In this paper, we present \textbf{CookVoice}, a unified framework for multimodal, multi-style, and multi-task human voice generation. By decomposing human voice into content, prosody, and style, CookVoice supports a wide range of speech and singing voice generation tasks within a single model, including TTS, TTSV, style-controllable generation, voice mimicry, and voice conversion. To enable fine-grained controllability, we design a flexible alignment strategy that expands different control signals to the level of the acoustic frame, allowing text, reference voice, discrete prosody, and continuous $F_0$ contours to be combined in a unified generation process. Experimental results show that CookVoice achieves comparable generation quality to existing task-specific baselines while providing stronger style and prosody controllability, smaller model size and efficient inference speed.

\section{Limitation}
CookVoice shows potential for unified and controllable human voice generation, several limitations remain. First, due to resource constraints, CookVoice has not yet been scaled up. The current model only use the DiT-S version of diffusion transformer~($43.51$M parameters) and is trained on $168$ hours of data, which is still much smaller than large-scale voice generation systems. The scaling potential of CookVoice has not been fully explored. Another limitation lies in CookVoice mainly focuses on human voice generation. Its potential applicability to broader audio generation domains, such as music, instrumental sound, and general audio generation, has not yet been investigated.

\bibliographystyle{IEEEtran}
\bibliography{reference}
\appendix
\section{Related Work}
\textbf{Autoregressive voice generation.}
Recent advances in neural voice generation have significantly improved the naturalness and expressiveness of synthesized speech. Representative systems such as CosyVoice~\cite{du2024cosyvoice} and IndexTTS~\cite{deng2025indextts} adopt autoregressive~(AR) generation mechanisms to model acoustic token sequences. Using token-by-token prediction and large-scale training, these models can generate expressive high-quality speech. However, AR decoding also introduces several limitations for controllability. Since duration, rhythm, and prosody are implicitly determined by the generated token sequence, the generation process is difficult to interpret and hard to control at a fine temporal granularity. This becomes especially problematic when the target task requires explicit duration or pitch control, such as singing voice generation, prosody mimicry, and frame-level voice editing. In addition, AR generation usually suffers from slower inference due to its sequential decoding process.

\textbf{NAR and controllable speech generation.}
NAR methods improve inference efficiency and temporal stability by introducing explicit or semi-explicit alignment mechanisms. For example, FastSpeech-style models rely on duration prediction and phoneme-level expansion to generate speech in parallel, while recent flow-matching or diffusion-based systems further improve acoustic quality through continuous generative modeling~\cite{chen2025f5}. These methods provide better control over phoneme-level timing than AR models. However, most existing NAR speech generation systems are still designed for specific tasks, such as TTS, expressive TTS, or voice cloning. Their control interfaces are often limited to either text prompts, speaker references, or phoneme-level durations, making it difficult to flexibly combine content, style,  prosody, and singing-related controls under a unified framework.

\textbf{Singing voice generation.}
Singing voice generation requires more explicit temporal and melodic control than ordinary speech synthesis. Traditional singing voice synthesis systems, such as DiffSinger~\cite{liu2022diffsinger}, usually rely on musical scores and note-level conditions to generate singing voices. More recent work introduces style control into singing voice generation. For example, StyleSinger~\cite{zhang2024stylesinger} improves controllability by incorporating singing style information. However, such systems often assume a predefined relationship between lyrics, phonemes, and notes. In particular, binding notes directly with phonemes simplifies alignment, but it may not always reflect the flexible temporal structure of real singing, where one phoneme may span multiple notes or one note may correspond to multiple phonetic units. This type of restricted alignment improves tractability but limits flexible control in more general singing voice generation scenarios.

\textbf{Unified voice generation.}
Recent unified voice generation models attempt to cover multiple voice generation tasks using a single model. For example, Vevo-style systems~\cite{vevo,zhang2026bm} move toward unified speech and singing voice modeling. However, many of these systems still rely on AR decoding or use global reference conditions, such as utterance-level or sentence-level prosody references. While these designs are effective for general voice generation, they provide limited direct control over fine-grained temporal details. In particular, when content, style, notes, lexical tones, stresses, and continuous $F_0$ contours are jointly involved, existing unified models still struggle to provide flexible frame-level controllability.

\section{Evaluation Metrics}
\label{app:metric_details}

This section provides detailed definitions of the evaluation metrics used in our experiments.

\noindent \textbf{Word Error Rate (WER).}
WER measures word-level transcription errors between the generated audio and the ground-truth text. We first transcribe the generated audio using Whisper~\cite{radford2023robust}, and then compute the edit distance between the predicted word sequence and the reference word sequence:
\begin{equation}
    \mathrm{WER} = \frac{S + D + I}{N},
\end{equation}
where $S$, $D$, and $I$ denote the number of substitutions, deletions, and insertions, respectively, and $N$ is the number of words in the reference text. Lower WER has better intelligibility.

\noindent \textbf{Phoneme Error Rate (PhoER).}
PhoER measures content accuracy at the phoneme level. We convert both the recognized text and the ground-truth text into phoneme sequences using the same G2P module, and then compute the normalized edit distance:
\begin{equation}
    \mathrm{PhoER} = \frac{S_p + D_p + I_p}{N_p},
\end{equation}
where $S_p$, $D_p$, and $I_p$ denote phoneme-level substitutions, deletions, and insertions, and $N_p$ is the number of phonemes in the reference sequence. A lower PhoER indicates more accurate pronunciation and content preservation.

\noindent \textbf{Prosody Error Rate (ProER).}
ProER measures the error rate of discrete prosody tokens, such as Mandarin tones, English stresses, or note-related prosody labels. Similar to PhoER, we compute the normalized edit distance between the predicted and reference prosody-token sequences:
\begin{equation}
    \mathrm{ProER} = \frac{S_r + D_r + I_r}{N_r},
\end{equation}
where $S_r$, $D_r$, and $I_r$ denote substitution, deletion, and insertion errors in the prosody-token sequence, and $N_r$ is the number of reference prosody tokens. A lower ProER indicates better preservation of discrete prosodic information.

\noindent \textbf{Style Similarity (S-SIM).}
S-SIM evaluates whether the generated voice matches the target style. We extract style embeddings from the generated audio and the target reference using a pre-trained style encoder. The cosine similarity between the two embeddings is computed as:
\begin{equation}
    \mathrm{S\text{-}SIM} = 
    \frac{\mathbf{s}_{gen}^{\top}\mathbf{s}_{ref}}
    {\lVert \mathbf{s}_{gen} \rVert_2 \lVert \mathbf{s}_{ref} \rVert_2},
\end{equation}
where $\mathbf{s}_{gen}$ and $\mathbf{s}_{ref}$ are the style embeddings of the generated and reference voices, respectively. A higher S-SIM indicates better style preservation.

\noindent \textbf{F0 Root Mean Square Error (F0-RMSE).}
F0-RMSE measures the absolute pitch difference between the generated and target F0 contours. We compute F0-RMSE over voiced frames where both generated and reference F0 values are valid:
\begin{equation}
    \mathrm{F0\text{-}RMSE} =
    \sqrt{
    \frac{1}{T_v}
    \sum_{t=1}^{T_v}
    \left( f^{gen}_t - f^{ref}_t \right)^2
    },
\end{equation}
where $f^{gen}_t$ and $f^{ref}_t$ are the generated and reference F0 values at frame $t$, and $T_v$ is the number of valid voiced frames. A lower F0-RMSE indicates smaller absolute pitch deviation.

\noindent \textbf{F0 Pearson Correlation (F0-CORR).}
F0-CORR measures whether the generated pitch contour follows the same trend as the reference contour:
\begin{equation}
    \mathrm{F0\text{-}CORR} =
    \frac{
    \sum_{t=1}^{T_v}
    (f^{gen}_t - \bar{f}^{gen})
    (f^{ref}_t - \bar{f}^{ref})
    }
    {
    \sqrt{\sum_{t=1}^{T_v}(f^{gen}_t - \bar{f}^{gen})^2}
    \sqrt{\sum_{t=1}^{T_v}(f^{ref}_t - \bar{f}^{ref})^2}
    },
\end{equation}
where $\bar{f}^{gen}$ and $\bar{f}^{ref}$ are the mean F0 values of the generated and reference contours. A higher F0-CORR indicates better preservation of pitch movement and prosodic trend.

\noindent \textbf{Mean Opinion Score (MOS).}
MOS is used to evaluate perceptual naturalness and overall audio quality. Human listeners rate each generated sample on a scale from 1 to 5, where 1 indicates poor quality and 5 indicates excellent quality. The final MOS is the average score across all listeners and samples.

\noindent \textbf{Melody Comparative MOS (M-CMOS).}
M-CMOS is used for singing voice evaluation. Listeners compare generated singing samples and judge which one better follows the target melody. This metric evaluates melody consistency from human perception, complementing objective F0-based metrics.

\section{Signal Impact Analysis}\label{sec:analysis}
We analyze the impact of the number of ODE solver steps under different sources of style and prosody control. Specifically, we compare text-based and voice-based style control, as well as discrete and continuous prosody control, on both TTS and TTSV tasks. We evaluate three aspects of generation quality: style similarity, prosody fidelity, and intelligibility.

\subsection{Style similarity}
\begin{figure}[htbp]
    \centering
    \begin{subfigure}{0.49\linewidth}
        \centering
        \includegraphics[width=\linewidth]{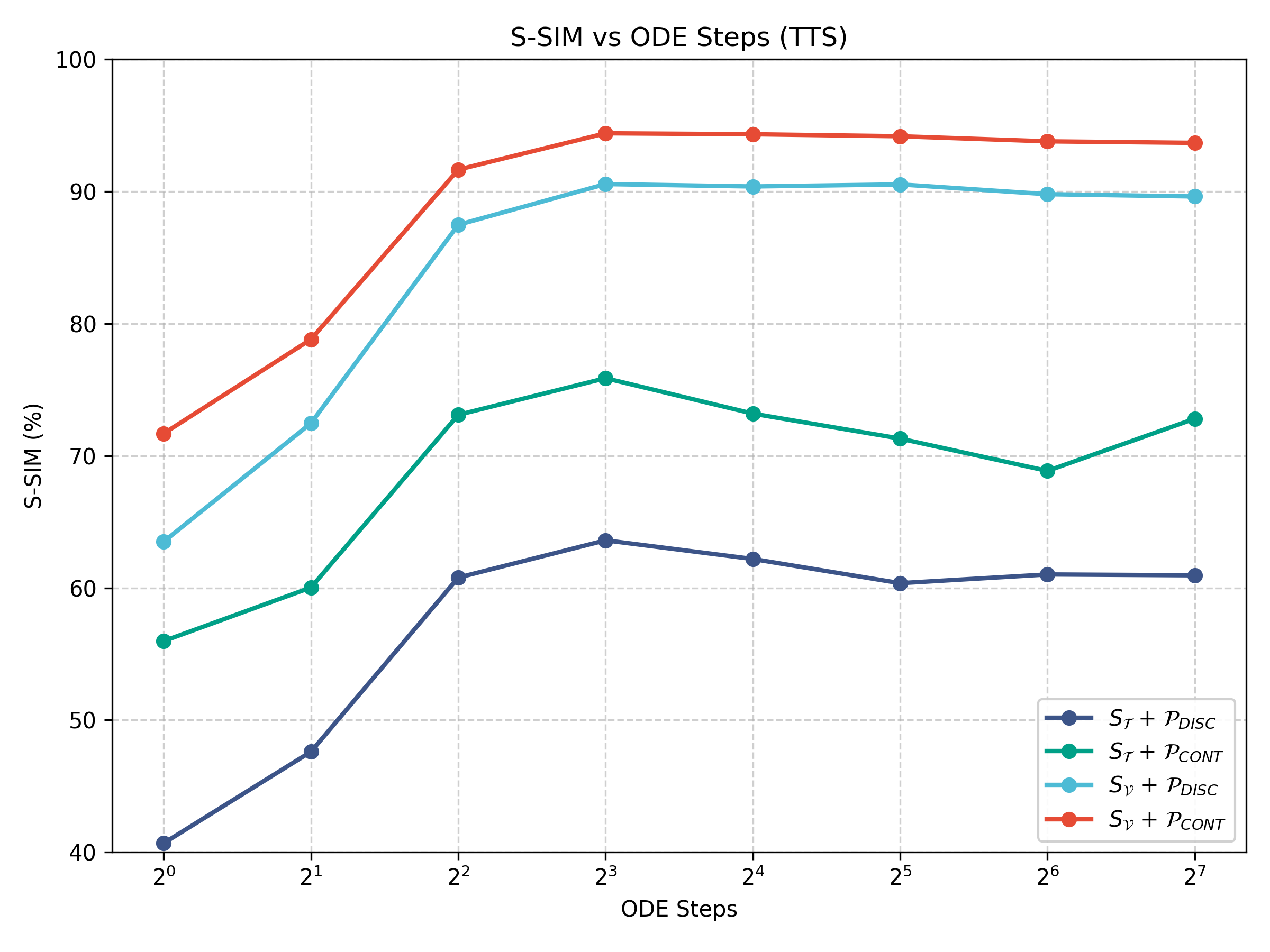}
        \caption{TTS}
        \label{fig:ode_ssim_tts}
    \end{subfigure}
    \hfill
    \begin{subfigure}{0.49\linewidth}
        \centering
        \includegraphics[width=\linewidth]{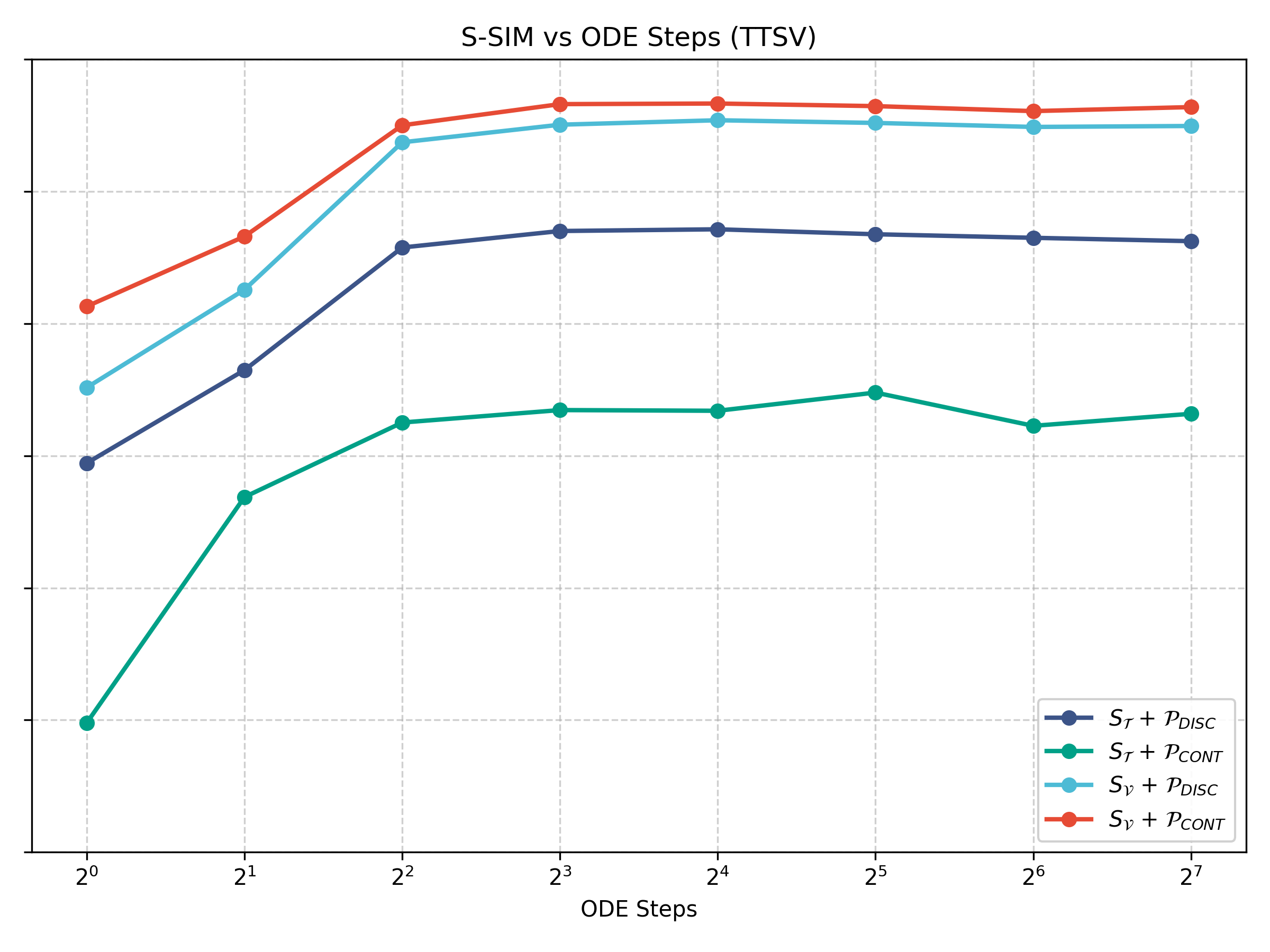}
        \caption{TTSV}
        \label{fig:ode_ssim_ttsv}
    \end{subfigure}
    \caption{Style similarity under different numbers of ODE steps for TTS and TTSV.}
    \label{fig:ode_ssim}
\end{figure}

As shown in Fig.~\ref{fig:ode_ssim}, style similarity generally increases as the number of ODE steps increases, and gradually converges around 4--8 steps. This suggests that most style-related information is formed in the early part of the generation trajectory, while further increasing the number of steps brings diminishing returns. We also observe that reference-voice-based style control consistently outperforms text-based style control. This is expected because reference voices contain richer speaker identity and paralinguistic information than text style descriptions. In addition, continuous prosody control generally achieves higher style similarity than discrete prosody tokens, indicating that dense frame-level control signals provide more detailed guidance during generation.

\subsection{Prosody fidelity}
\begin{figure}[htbp]
    \centering
    \begin{subfigure}{0.49\linewidth}
        \centering
        \includegraphics[width=\linewidth]{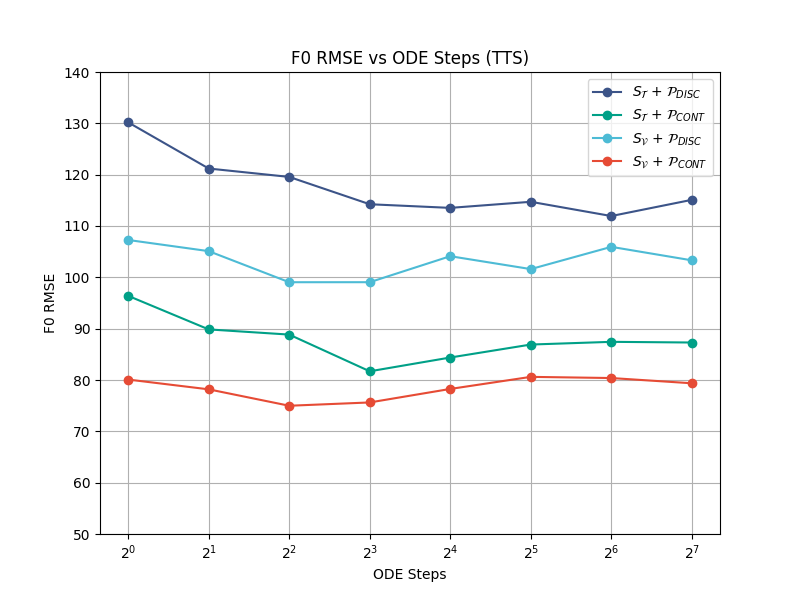}
        \caption{F0-RMSE on TTS}
        \label{fig:ode_rmse_tts}
    \end{subfigure}
    \hfill
    \begin{subfigure}{0.49\linewidth}
        \centering
        \includegraphics[width=\linewidth]{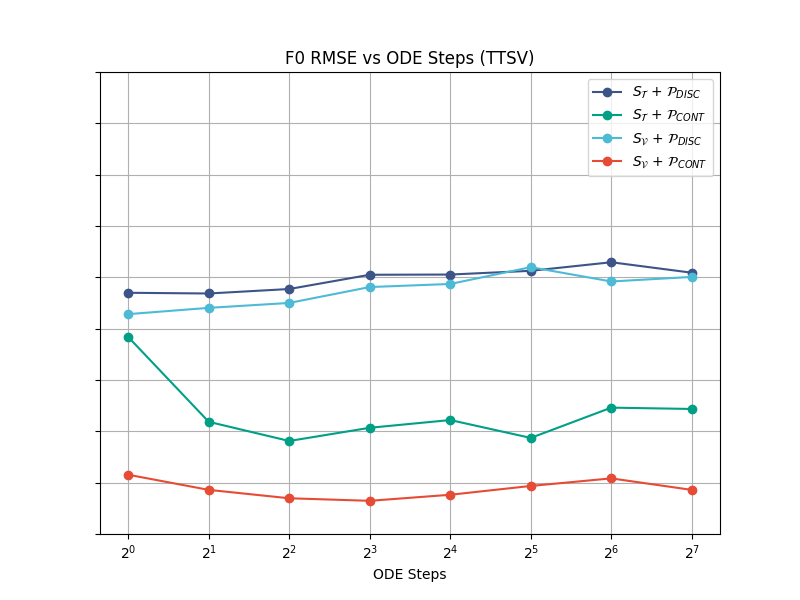}
        \caption{F0-RMSE on TTSV}
        \label{fig:ode_rmse_ttsv}
    \end{subfigure}

    \vspace{0.5em}

    \begin{subfigure}{0.49\linewidth}
        \centering
        \includegraphics[width=\linewidth]{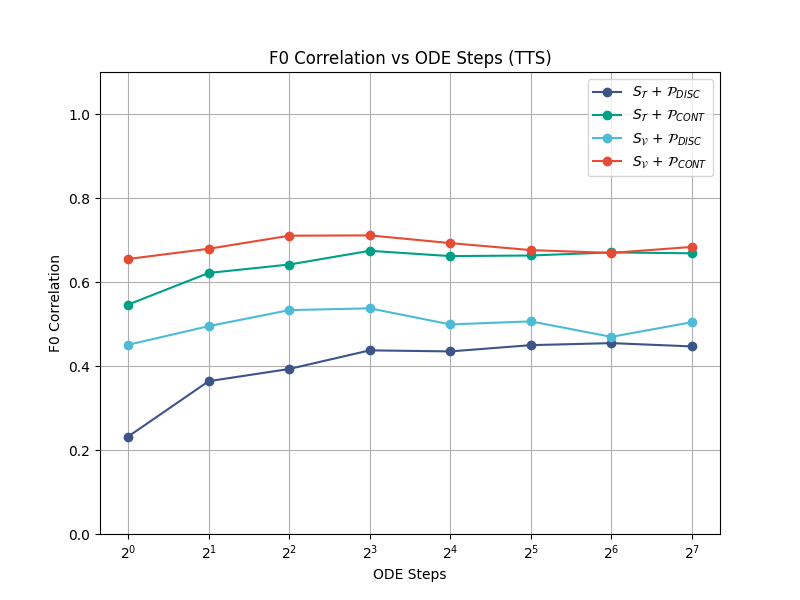}
        \caption{F0-Correlation on TTS}
        \label{fig:ode_corr_tts}
    \end{subfigure}
    \hfill
    \begin{subfigure}{0.49\linewidth}
        \centering
        \includegraphics[width=\linewidth]{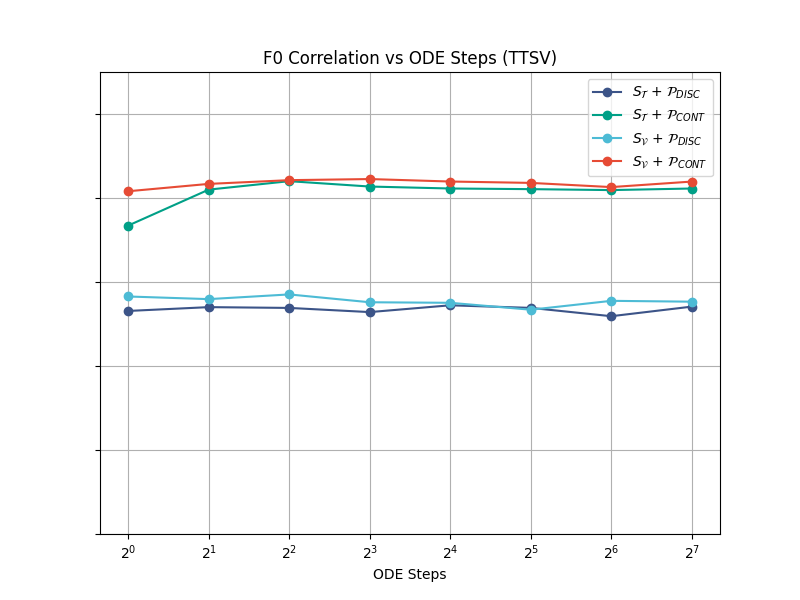}
        \caption{F0-Correlation on TTSV}
        \label{fig:ode_corr_ttsv}
    \end{subfigure}
    \caption{Prosody fidelity under different numbers of ODE steps. F0-RMSE measures absolute pitch deviation, while F0-Correlation measures the consistency of the generated F0 contour with the target contour.}
    \label{fig:ode_prosody}
\end{figure}

For prosody, we evaluate F0-RMSE and F0-Correlation, as shown in Fig.~\ref{fig:ode_prosody}. F0-RMSE measures the absolute pitch deviation from the ground-truth contour, while F0-Correlation measures whether the generated pitch follows the overall trend of the target F0 trajectory. For F0-RMSE, the error generally decreases as the number of ODE steps increases, with voice-based style and continuous prosody control achieving better performance. This is also reasonable because the reference voice directly provides acoustic and pitch-related cues, while continuous F0 signals preserve more fine-grained pitch information than discrete tokens.

A similar trend can be observed for F0-Correlation, where the correlation improves with more ODE steps and converges around 8 steps. However, the improvement for discrete prosody control is relatively limited, especially in the TTSV setting. One possible reason is that singing voice often follows a more constrained and stable melodic structure compared with speech. Therefore, discrete prosody tokens may already capture the coarse pitch movement, and increasing the number of ODE steps mainly improves acoustic details rather than substantially changing the global F0 correlation.

\subsection{Intelligibility}
\begin{figure}[htbp]
    \centering
    \begin{subfigure}{0.49\linewidth}
        \centering
        \includegraphics[width=\linewidth]{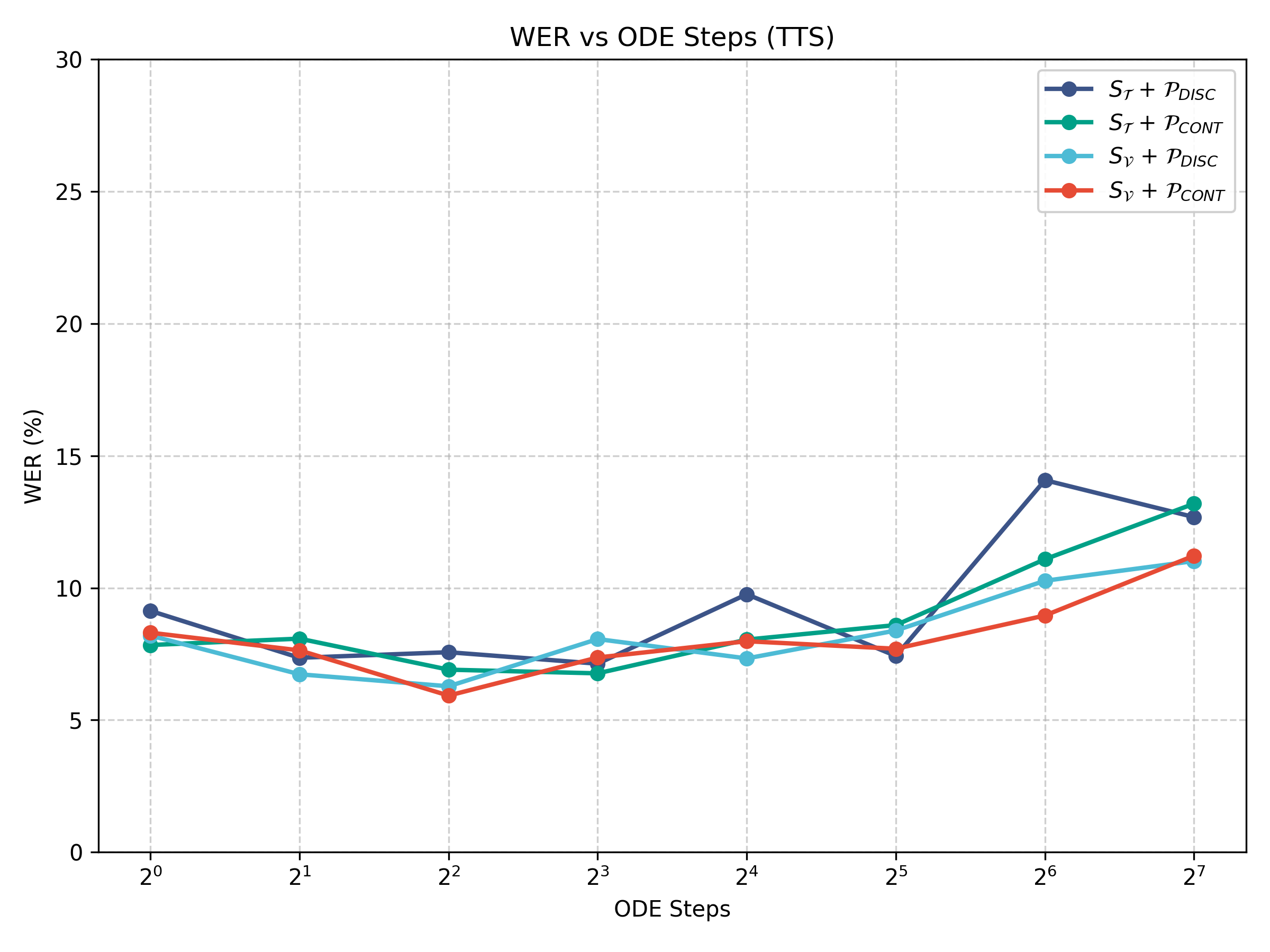}
        \caption{WER on TTS}
        \label{fig:ode_wer_tts}
    \end{subfigure}
    \hfill
    \begin{subfigure}{0.49\linewidth}
        \centering
        \includegraphics[width=\linewidth]{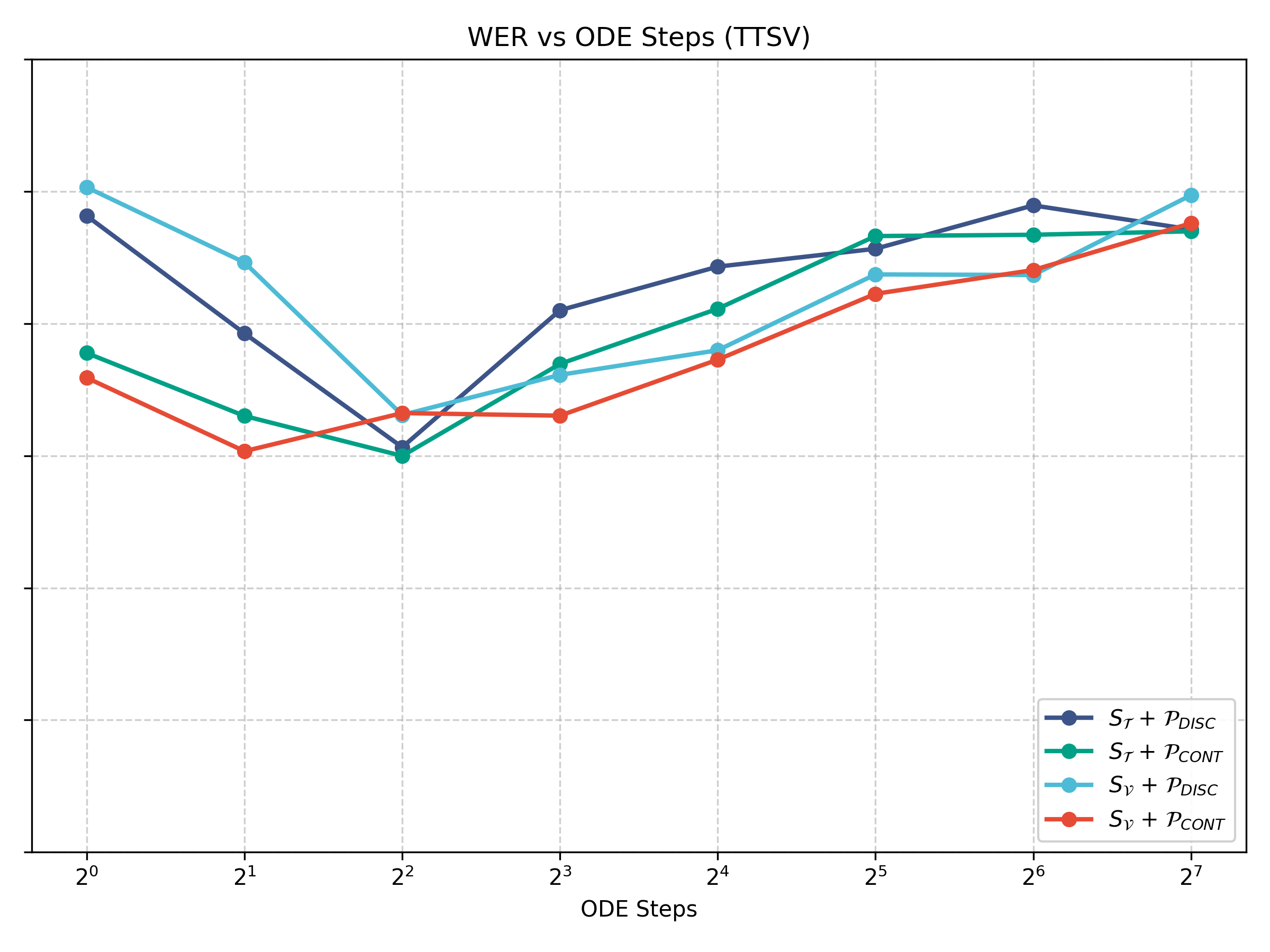}
        \caption{WER on TTSV}
        \label{fig:ode_wer_ttsv}
    \end{subfigure}

    \vspace{0.5em}

    \begin{subfigure}{0.49\linewidth}
        \centering
        \includegraphics[width=\linewidth]{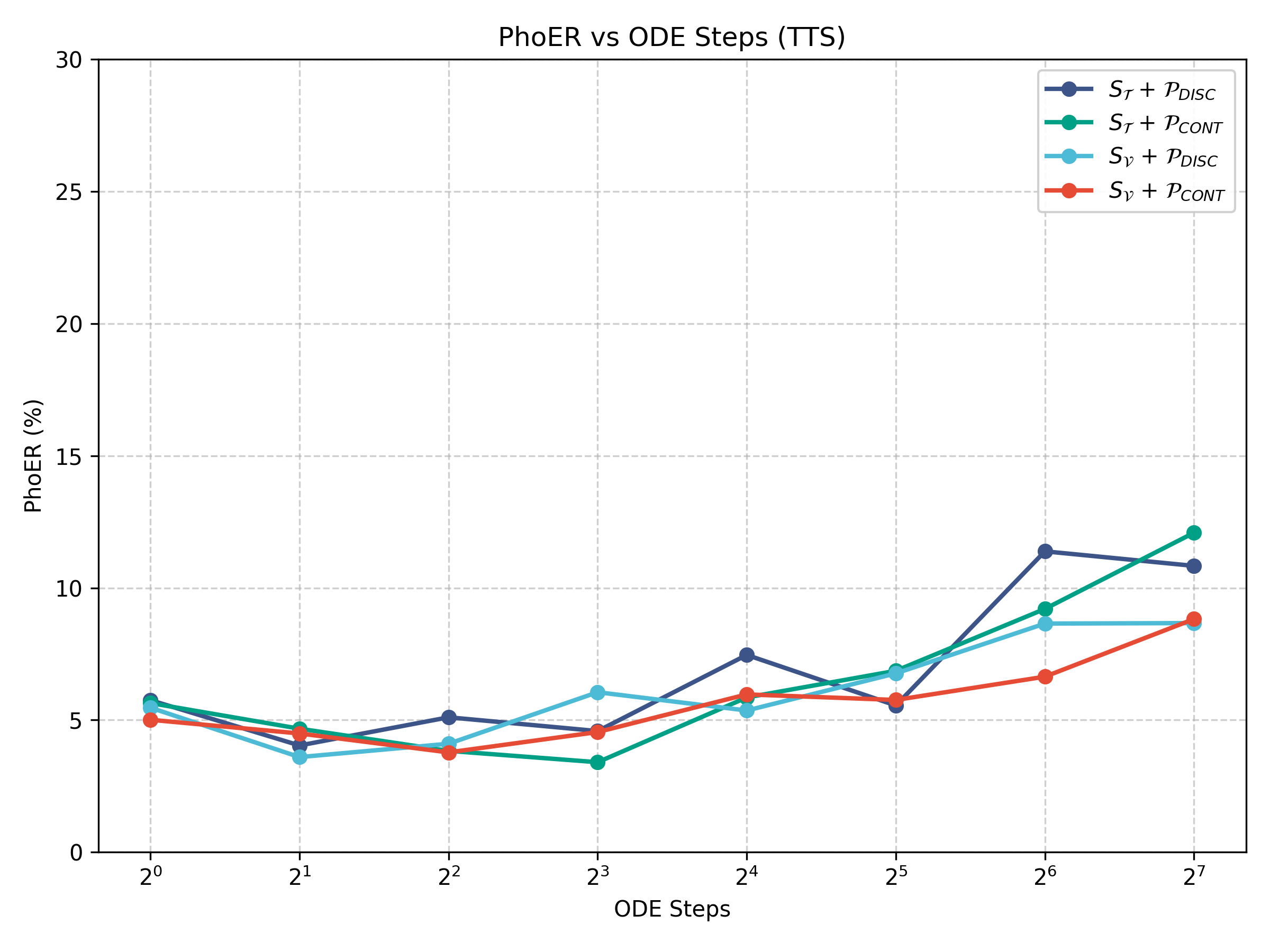}
        \caption{PhoER on TTS}
        \label{fig:ode_phoer_tts}
    \end{subfigure}
    \hfill
    \begin{subfigure}{0.49\linewidth}
        \centering
        \includegraphics[width=\linewidth]{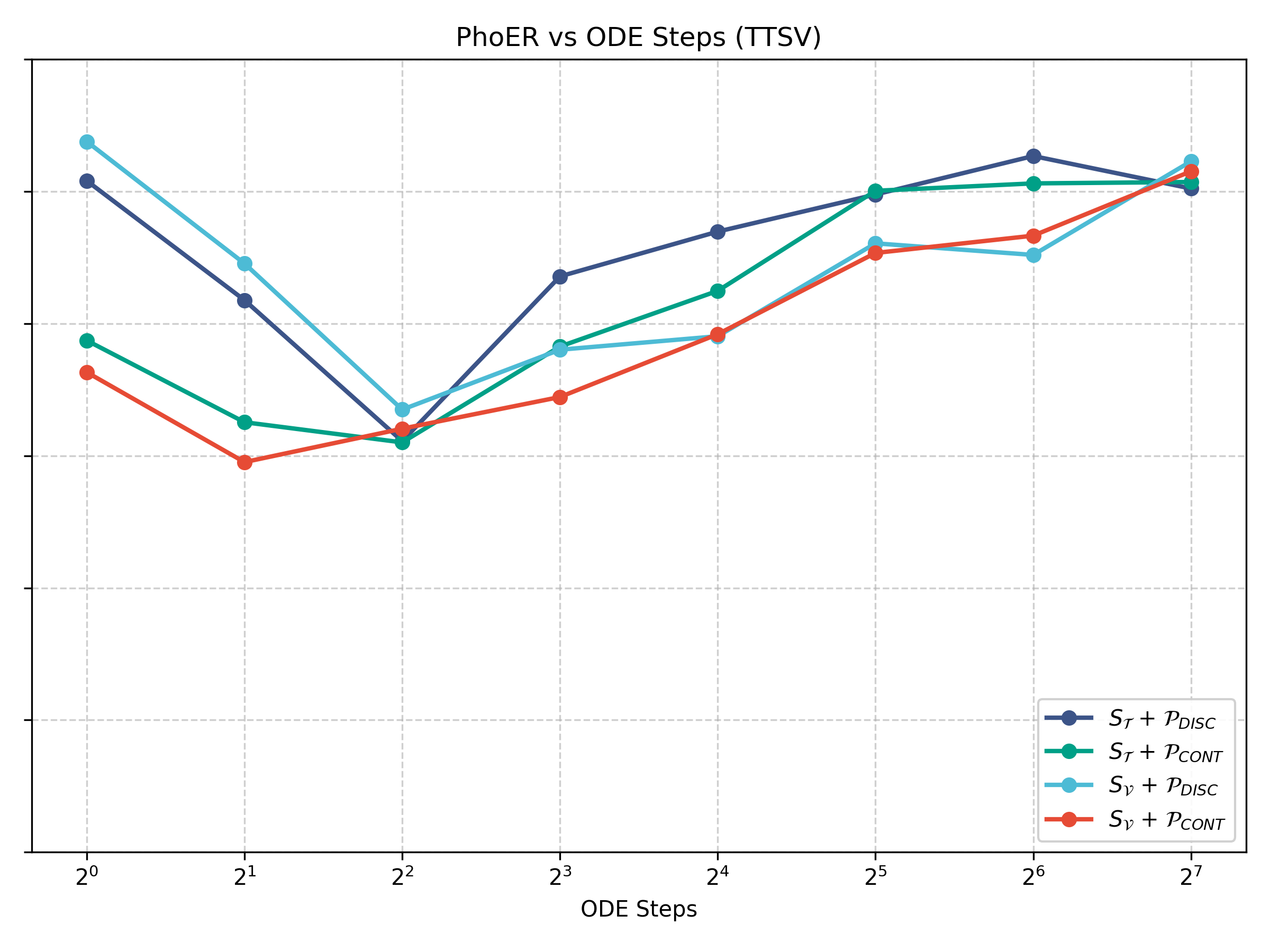}
        \caption{PhoER on TTSV}
        \label{fig:ode_phoer_ttsv}
    \end{subfigure}

    \vspace{0.5em}

    \begin{subfigure}{0.49\linewidth}
        \centering
        \includegraphics[width=\linewidth]{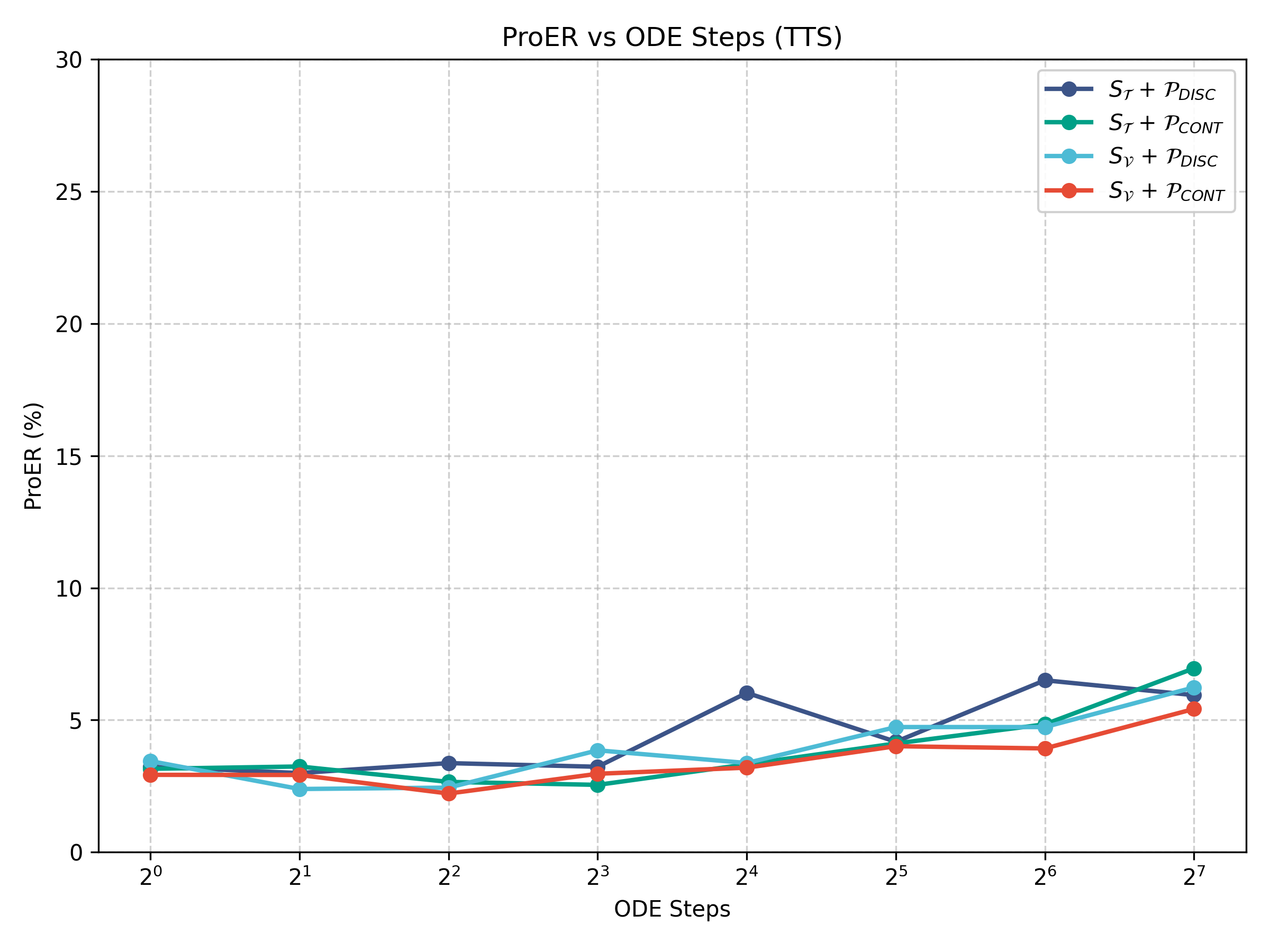}
        \caption{ProER on TTS}
        \label{fig:ode_proer_tts}
    \end{subfigure}
    \hfill
    \begin{subfigure}{0.49\linewidth}
        \centering
        \includegraphics[width=\linewidth]{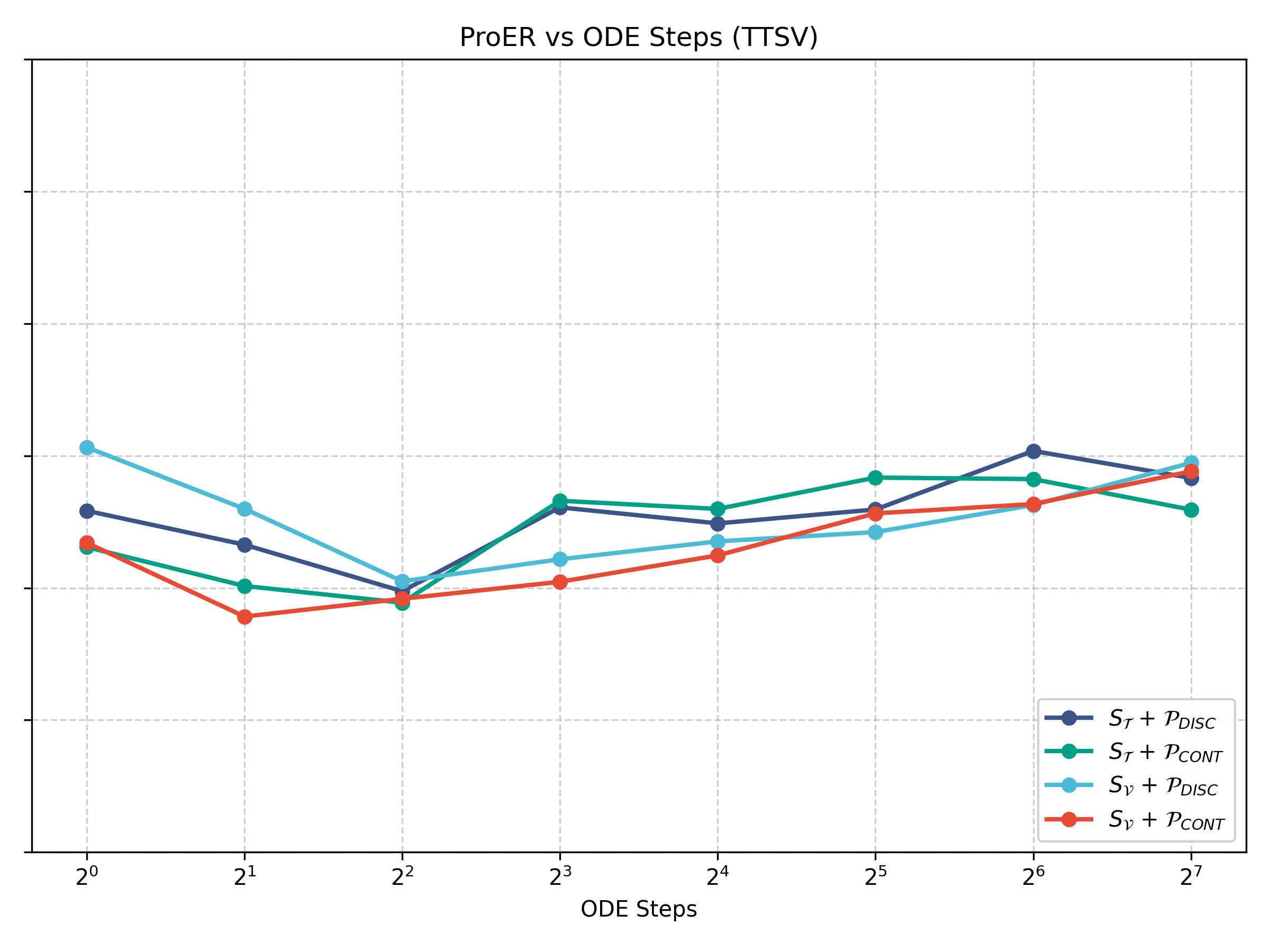}
        \caption{ProER on TTSV}
        \label{fig:ode_proer_ttsv}
    \end{subfigure}
    \caption{Intelligibility under different numbers of ODE steps for TTS and TTSV. WER, PhoER, and ProER are used to evaluate word-level, phoneme-level, and prosody-token-level error rates, respectively.}
    \label{fig:ode_intelligibility}
\end{figure}

The intelligibility results are shown in Fig.~\ref{fig:ode_intelligibility}. Different from style similarity and prosody fidelity, WER, PhoER, and ProER first decrease and reach their best performance at 4 ODE steps, but tend to increase when more ODE steps are used. This indicates that using too many ODE steps may over-refine acoustic, style, or prosodic details at the cost of linguistic clarity. In other words, a larger number of ODE steps does not always lead to better content preservation.

We also observe that TTSV generally has higher error rates than TTS. This may be partly attributed to the evaluation model, since Whisper~\cite{radford2023robust} is mainly trained and optimized for speech recognition rather than singing voice recognition. Moreover, singing voice naturally contains longer phoneme durations, stronger pitch variation, and melody-driven pronunciation changes, which can make recognition more difficult.

Overall, these results suggest that different generation aspects prefer different numbers of ODE steps. Style similarity and prosody fidelity benefit from increasing ODE steps and usually converge around 4--8 steps, while intelligibility achieves the best trade-off at around 4 steps. Therefore, a small number of ODE steps provides a better balance between controllability, acoustic quality, and content preservation.

\begin{figure}[htbp]
    \centering
    \begin{subfigure}{0.49\linewidth}
        \centering
        \includegraphics[width=\linewidth]{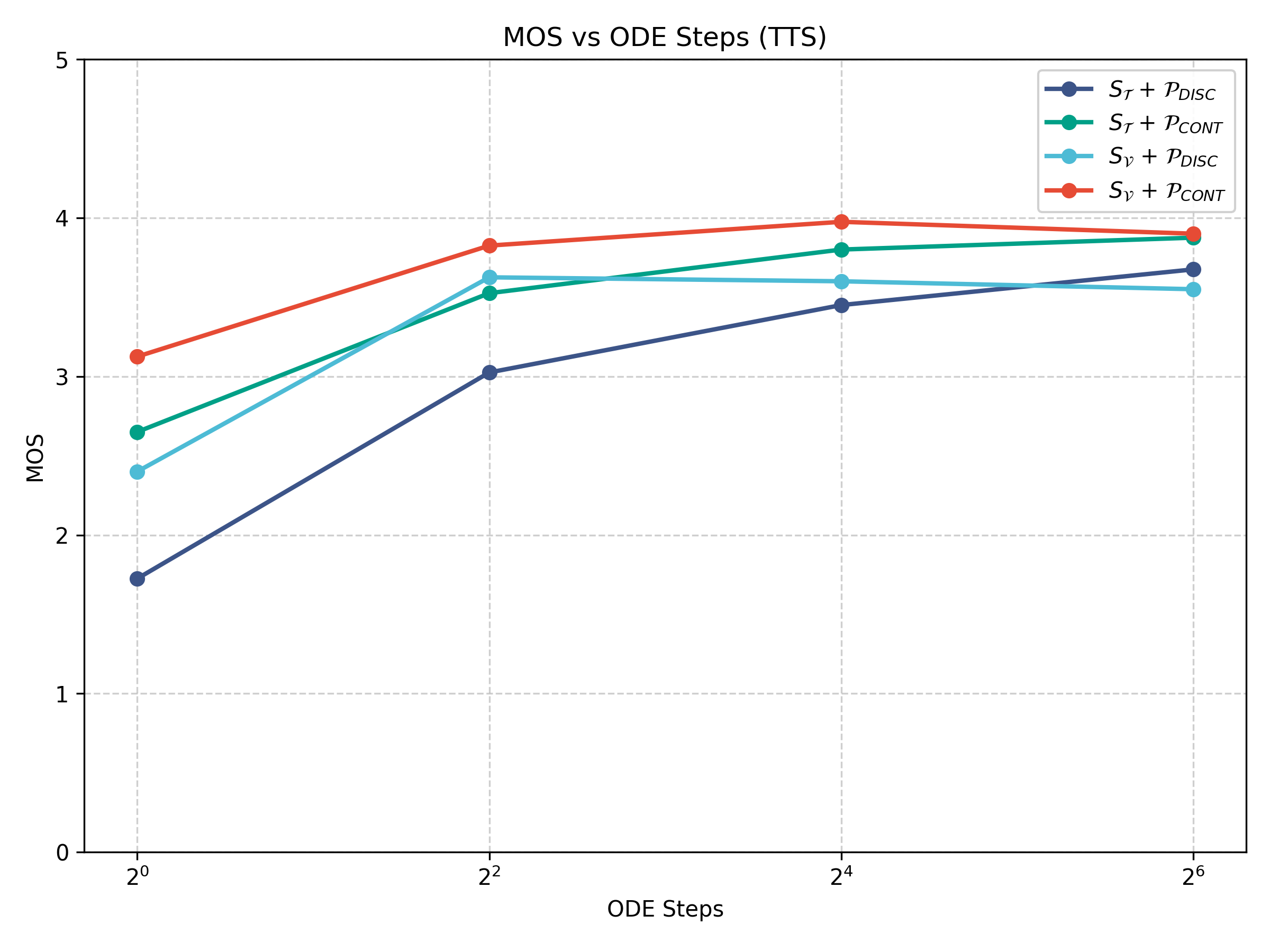}
        \caption{MOS on TTS}
        \label{fig:ode_mos_tts}
    \end{subfigure}
    \hfill
    \begin{subfigure}{0.49\linewidth}
        \centering
        \includegraphics[width=\linewidth]{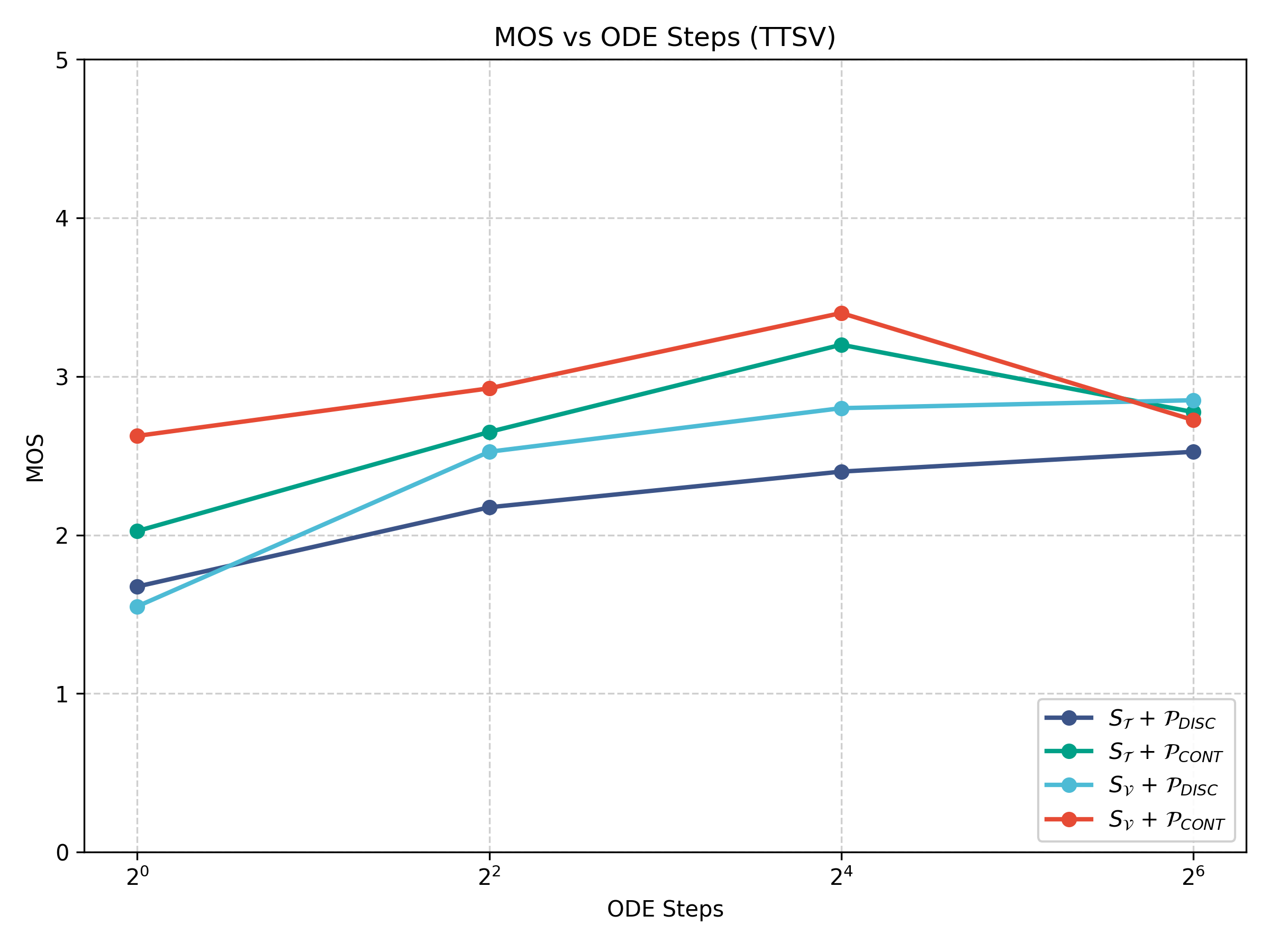}
        \caption{MOS on TTSV}
        \label{fig:ode_mos_ttsv}
    \end{subfigure}

    \vspace{0.5em}

    \begin{subfigure}{0.49\linewidth}
        \centering
        \includegraphics[width=\linewidth]{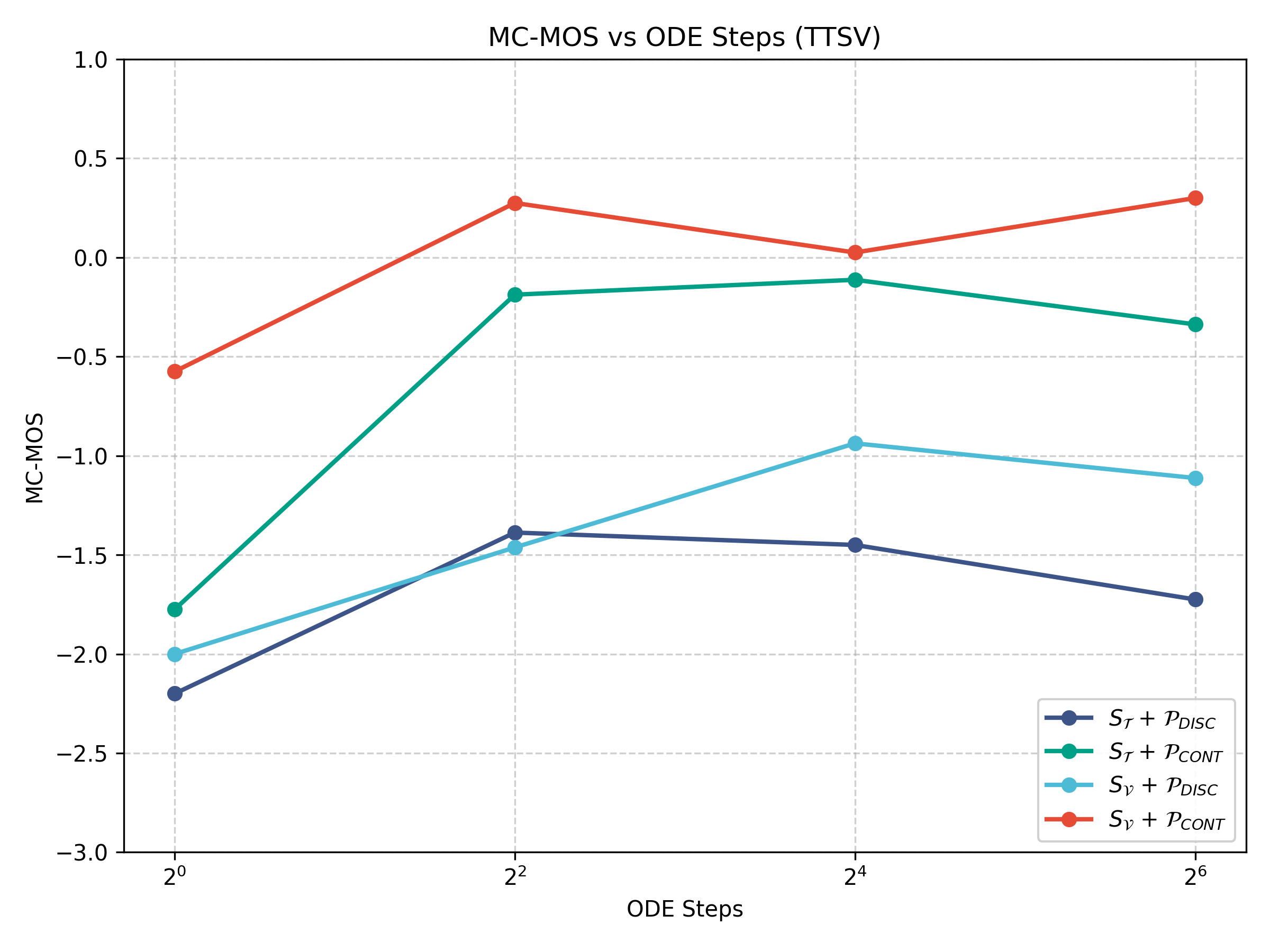}
        \caption{MC-MOS on TTSV}
        \label{fig:ode_mcmos_ttsv}
    \end{subfigure}
    
    \caption{Human subjective listening test evaluation under different numbers of ODE steps. MOS measures overall naturalness, MC-MOS measures the melody accuracy of the generated singing voice.}
    \label{fig:ode_mos}
\end{figure}

\section{Task Definition}\label{app:definition}
CookVoice is designed as a unified framework for speech and singing voice generation. Instead of building separate models for different tasks, we formulate a broad range of generation, conversion, editing, and mimicry tasks under a shared conditional generation function $\mathcal{G}(\cdot)$. The key difference between tasks lies in the combination of control signals, including linguistic content, style conditions, lexical or musical prosody, and frame-level pitch contours.

Specifically, $X$ represents the phoneme derived from text or lyric to be generated, while $S_\mathcal{T}$ and $S_\mathcal{V}$ denote style information provided by text descriptions and reference voices, respectively. Prosodic control can be provided at different granularities: $\mathcal{P}_{lex}$ represents discrete lexical prosody such as tones or stresses, $\mathcal{P}_{note}$ represents musical note-level conditions for singing voice generation, and $F_0$ provides continuous frame-level pitch control. By flexibly combining these conditions, CookVoice can support multiple tasks within a single model, as summarized in Table~\ref{tab:task_definition}.

\begin{table*}[t!]
\centering
\small
\setlength{\tabcolsep}{4pt}
\caption{Summary of generation tasks and control settings supported by CookVoice. $X$ denotes phoneme/lyric content, $S_\mathcal{T}$ and $S_\mathcal{V}$ denote text-based and voice-based style conditions, $\mathcal{P}_{lex}$ denotes lexical prosody such as tones or stresses, $\mathcal{P}_{note}$ denotes musical note conditions, and $F_0$ denotes F0 contour.}
\label{tab:task_definition}
\begin{tabular}{p{0.25\linewidth} p{0.08\linewidth} p{0.17\linewidth} p{0.43\linewidth}}
\toprule
\textbf{Task Name} & \textbf{Abbr.} & \textbf{Formulation} & \textbf{Description} \\
\midrule

Text-to-Speech 
& TTS 
& $\mathcal{G}(X,\mathcal{P}_{lex})$ 
& Generate a speech waveform from text content and lexical prosody. \\

Text Style-Controllable TTS 
& T-SC-TTS 
& $\mathcal{G}(S_\mathcal{T},X,\mathcal{P}_{lex})$ 
& Generate speech conditioned on a natural language style description, such as emotion, age, gender, or speaking manner. \\

Voice Style-Controllable TTS 
& V-SC-TTS 
& $\mathcal{G}(S_\mathcal{V},X,\mathcal{P}_{lex})$ 
& Generate speech that preserves the text content while mimicking the global style or timbre of a reference voice. \\

Prosody-Controllable TTS 
& PC-TTS 
& $\mathcal{G}(S,X,F_0)$ 
& Generate speech with explicit frame-level prosody control from a reference $F_0$ contour while keeping the target content and style. \\

\midrule

Text-to-Singing Voice 
& TTSV 
& $\mathcal{G}(X,\mathcal{P}_{note})$ 
& Generate a singing voice from lyrics and musical notes. \\

Text Style-Controllable TTSV 
& T-SC-TTSV 
& $\mathcal{G}(S_\mathcal{T},X,\mathcal{P}_{note})$ 
& Generate singing voice from lyrics and notes while controlling singing style using a text prompt. \\

Voice Style-Controllable TTSV 
& V-SC-TTSV 
& $\mathcal{G}(S_\mathcal{V},X,\mathcal{P}_{note})$ 
& Generate singing voice from lyrics and notes while transferring the style or timbre from a reference voice. \\

Prosody-Controllable TTSV 
& PC-TTSV 
& $\mathcal{G}(S,X,\mathcal{P}_{F_0})$ 
& Generate singing voice by following a continuous frame-level $F_0$ contour, enabling fine-grained melody or prosody mimicry. \\

\midrule

Voice Mimicry 
& VM 
& $\mathcal{G}(S_\mathcal{V},X,\mathcal{P}_{F_0})$ 
& Generate speech with new content while jointly mimicking the reference voice style and detailed prosodic trajectory. \\

Voice Conversion 
& VC 
& $\mathcal{G}(S_\mathcal{V}^{tar},X^{src},\mathcal{P}_{F_0}^{src})$ 
& Convert a source speech into a target voice style while preserving the source linguistic content and prosody. \\

Singing Voice Conversion 
& SVC 
& $\mathcal{G}(S_\mathcal{V}^{tar},X^{src},\mathcal{P}_{F_0}^{src})$ 
& Convert a source singing voice into a target singing style while preserving lyrics and melodic contour. \\

\midrule

Speech Editing 
& SE 
& $\mathcal{G}(S,X^{edit},\mathcal{P}_{F_0})$ 
& Edit the linguistic content or style of speech while maintaining the remaining prosodic structure. \\

Singing Voice Editing 
& SVE 
& $\mathcal{G}(S,X^{edit},\mathcal{P}_{note/F_0})$ 
& Edit lyrics, style, or melody-related controls in singing voice generation. \\

Sketch-to-Voice 
& STV 
& $\mathcal{G}(S,X,\mathcal{P}_{F_0}^{sketch})$ 
& Generate speech or singing voice from a user-specified coarse pitch sketch or manually controlled $F_0$ trajectory. \\

Humming-To-Voice
& HTV
& $\mathcal{G}(S,X,\mathcal{P}_{F_0}^{humm})$
& Generate speech or singing voice from a user-specified humming $F_0$ contour. \\
\bottomrule
\end{tabular}

\end{table*}

\section{Terminology}\label{app:terminology}
\label{sec:terminology}

To make the paper accessible to readers without a background in linguistics, speech processing, or music, we summarize the key terms used in CookVoice in Table~\ref{tab:terminology}. These terms cover linguistic content, prosody, music-related control signals, style control, and evaluation metrics.

\begin{table*}[t!]
\centering
\small
\setlength{\tabcolsep}{4pt}
\caption{Terminology used in CookVoice. The table provides simplified explanations of linguistic, musical, control, and evaluation terms for non-expert readers.}
\label{tab:terminology}
\begin{tabular}{p{0.18\linewidth} p{0.16\linewidth} p{0.58\linewidth}}
\toprule
\textbf{Term} & \textbf{Category} & \textbf{Meaning in CookVoice} \\
\midrule

Human Voice Generation (HVG)
& General task
& A broad task family that generates human speech or singing voice from control signals such as text, lyrics, notes, style prompts, reference voices, or pitch contours. \\

Text-to-Speech (TTS)
& Speech task
& Generate a spoken voice waveform from text or phoneme content. \\

Text-to-Singing Voice (TTSV)
& Singing task
& Generate a singing voice waveform from lyrics and musical note information. \\

Voice Conversion (VC)
& Conversion task
& Convert a source speech voice into a target speaker or style while preserving the source content and prosody. \\

Singing Voice Conversion (SVC)
& Conversion task
& Convert a source singing voice into a target singing voice style while preserving lyrics and melody. \\

\midrule

Phoneme
& Linguistics
& The basic sound unit of speech. For example, a word is converted into a sequence of phonemes before being generated as speech or singing voice. \\

Grapheme-to-Phoneme (G2P)
& Linguistics
& A text processing module that converts written text or lyrics into phoneme sequences. \\

Lyrics
& Music / linguistics
& The textual content of a song. In CookVoice, lyrics are converted into phonemes and used as singing content. \\

Lexical prosody
& Linguistics
& Discrete pronunciation-related information attached to phonemes or words, such as Mandarin tones or English word stresses. \\

Tone
& Linguistics
& A pitch pattern that changes word meaning in tonal languages such as Mandarin Chinese. In CookVoice, tones are used as discrete prosody tokens for speech. \\

Stress
& Linguistics
& The emphasis placed on a syllable or word, often affecting loudness, duration, and pitch in English speech. \\

\midrule

Prosody
& Speech / music
& The expressive pattern of voice, including pitch, rhythm, duration, stress, and intonation. Prosody controls how the content is spoken or sung. \\

Discrete prosody
& Control signal
& Symbolic prosody information such as tones, stresses, or MIDI notes. It provides coarse but practical control without requiring reference audio. \\

Continuous prosody
& Control signal
& Frame-level acoustic prosody information, mainly represented by the continuous $F_0$ contour. It provides fine-grained pitch control. \\

$F_0$
& Acoustics
& Fundamental frequency, corresponding to the perceived pitch of the voice. Higher $F_0$ generally means a higher-pitched voice. \\

$F_0$ contour
& Acoustics
& A time-varying sequence of $F_0$ values that describes how pitch changes over time. CookVoice uses it for frame-level continuous prosody control. \\

Waveform
& Signal processing
& The raw audio signal represented as amplitude values over time. It is the final audio output that listeners hear. \\

Spectrogram
& Signal processing
& A time-frequency representation of audio. It shows how the frequency content of a waveform changes over time, and is commonly used as an intermediate acoustic representation for voice generation. \\

Frame
& Signal processing
& The smallest unit of slice in the spectrogram along temporal axis. A voice waveform is converted into a spectrogram with multiple frames along the temporal axis, and CookVoice aligns content, style, and prosody to these acoustic frames for fine-grained control. \\

Latent acoustic embedding
& Signal processing
& A compressed acoustic feature extracted from the spectrogram. CookVoice generates the target latent acoustic embedding, which is then converted back into a waveform by the decoder. \\

Frame-level alignment
& Model design
& A mechanism that expands different control signals to match the acoustic frame length, allowing content, style, and prosody to be combined at each time step. \\

Duration
& Timing
& The length of time a phoneme or note lasts. Duration determines how content is expanded to match the acoustic frames. \\

\bottomrule
\end{tabular}
\end{table*}

\begin{table*}[t!]
\ContinuedFloat
\centering
\small
\setlength{\tabcolsep}{4pt}
\caption{Terminology used in CookVoice. The table provides simplified explanations of linguistic, musical, control, and evaluation terms for non-expert readers. Continued.}
\begin{tabular}{p{0.18\linewidth} p{0.16\linewidth} p{0.58\linewidth}}
\toprule
\textbf{Term} & \textbf{Category} & \textbf{Meaning in CookVoice} \\
\midrule

MIDI note
& Music
& A symbolic musical pitch representation used in singing voice generation. It specifies which note should be sung. \\

Note duration
& Music
& The length of a musical note, usually derived from the score or beat information. It controls how long a note is sustained in singing. \\

Beat
& Music
& A unit of musical timing. In CookVoice, beat information is used to compute relative note and phoneme durations for singing voice generation. \\

Melody
& Music
& The ordered pitch movement of singing. It is mainly controlled by musical notes or continuous $F_0$ contours. \\

Humming
& Music / control
& A user-provided vocal melody without full lyrics. CookVoice can use its extracted $F_0$ contour as a prosody control signal. \\

Pitch sketch
& User control
& A manually specified or coarse pitch trajectory used to guide the generated speech or singing voice. \\

\midrule

Style
& Voice attribute
& Global paralinguistic characteristics of a voice, such as speaker identity, timbre, emotion, age, gender, and speaking or singing manner. \\

Timbre
& Voice attribute
& The tone color or acoustic identity of a voice. Two voices can sing the same note but sound different because of different timbres. \\

Text style prompt
& Style control
& A natural language description of the desired voice style, such as ``a female speaking happily''. \\

Reference voice
& Style control
& An example speech or singing voice used to provide target voice style, timbre, or speaker identity. \\

Text-based style
& Style control
& Style information extracted from a natural language prompt. It provides semantic-level style control. \\

Voice-based style
& Style control
& Style information extracted from a reference voice. It provides more direct acoustic style and timbre control. \\

\midrule

WER
& Evaluation metric
& Word Error Rate. It measures word-level content errors after transcribing generated audio with an ASR system. Lower is better. \\

PhoER
& Evaluation metric
& Phoneme Error Rate. It measures phoneme-level pronunciation or content errors. Lower is better. \\

ProER
& Evaluation metric
& Prosody Error Rate. It measures errors in discrete prosody tokens such as tones and stresses. Lower is better. \\

S-SIM
& Evaluation metric
& Style Similarity. It measures how similar the generated voice style is to the target style. Higher is better. \\

F0-RMSE
& Evaluation metric
& Root mean square error between generated and target $F_0$ contours. It measures absolute pitch deviation. Lower is better. \\

F0-CORR
& Evaluation metric
& Pearson correlation between generated and target $F_0$ contours. It measures whether the pitch trend is preserved. Higher is better. \\

MOS
& Subjective metric
& Mean Opinion Score. Human listeners rate the naturalness or quality of generated audio, usually on a 1--5 scale. \\

M-CMOS
& Subjective metric
& Melody Comparative MOS. Human listeners compare whether one singing sample follows the melody better than another. \\

\bottomrule
\end{tabular}
\end{table*}
\section{Supplementary Results}
\begin{table*}[htbp]
\centering
\caption{Intelligibility and prosodic evaluation comparison on Chinese and English subsets. Word Error Rate~(\textbf{WER}), Phoneme Error Rate~(\textbf{PhoER}), Prosody Error Rate~(\textbf{ProER}) $\downarrow$ indicates lower is better.}
\label{tab:intelligibility_results_lang}
\renewcommand{\arraystretch}{1.1}
\begin{tabular}{ll| ccc ccc}
\toprule
& & \multicolumn{3}{c}{\textbf{Chinese}} & \multicolumn{3}{c}{\textbf{English}} \\
\cmidrule(lr){3-5} \cmidrule(lr){6-8}
\textbf{Task} & \textbf{Model} & \textbf{WER} & \textbf{PhoER} & \textbf{ProER} & \textbf{WER} & \textbf{PhoER} & \textbf{ProER} \\
\midrule
\multirow{10}{*}{TTS} 
& CosyVoice~\cite{du2024cosyvoice} & 8.62 $\pm$ 10.60 & 3.48 $\pm$ 12.19 & 2.63 $\pm$ 7.80 & 4.46 $\pm$ 8.30 & 3.93 $\pm$ 8.35 & 2.65 $\pm$ 6.71 \\
& F5-TTS~\cite{chen2025f5} & 5.23 $\pm$ 7.45 & 0.47 $\pm$ 1.67 & 0.63 $\pm$ 1.78 & 2.34 $\pm$ 4.67 & 1.63 $\pm$ 3.32 & 0.98 $\pm$ 1.96 \\
& ParaStyleTTS~\cite{lou2025parastyletts} & 9.67 $\pm$ 8.25 & 4.63 $\pm$ 6.70 & 3.27 $\pm$ 6.54 & 5.21 $\pm$ 6.95 & 5.40 $\pm$ 7.42 & 1.82 $\pm$ 2.80 \\
& IndexTTS~\cite{deng2025indextts} & 4.16 $\pm$ 6.63 & 0.35 $\pm$ 1.60 & 0.35 $\pm$ 1.60 & 2.88 $\pm$ 6.04 & 2.04 $\pm$ 5.26 & 1.29 $\pm$ 4.07 \\
& Vevo2~\cite{zhang2026bm} & 6.35 $\pm$ 6.73 & 1.56 $\pm$ 2.93 & 1.17 $\pm$ 2.62 & 3.82 $\pm$ 6.47 & 2.72 $\pm$ 6.01 & 1.62 $\pm$ 4.33 \\
\cmidrule(lr){2-8}
& \multicolumn{7}{l}{CookVoice} \\
& \quad $S_{\mathcal{T}} + \mathcal{P}_{disc}$ &  9.48 $\pm$ 11.88 & 4.22 $\pm$ 10.75 & 3.79 $\pm$ 9.17 &6.07 $\pm$ 12.35 & 5.80 $\pm$ 12.43 & 3.05 $\pm$ 6.09 \\
& \quad $S_{\mathcal{T}} + \mathcal{P}_{cont}$ &  9.03 $\pm$ 10.98 & 3.07 $\pm$ 8.65 & 2.69 $\pm$ 6.74 &5.23 $\pm$ 8.76 & 4.44 $\pm$ 8.85 & 2.64 $\pm$ 5.20 \\
& \quad $S_{\mathcal{V}} + \mathcal{P}_{disc}$ &  8.66 $\pm$ 10.14 & 4.28 $\pm$ 9.66 & 3.13 $\pm$ 6.97 &4.41 $\pm$ 7.40 & 3.97 $\pm$ 7.21 & 1.90 $\pm$ 3.50 \\
& \quad $S_{\mathcal{V}} + \mathcal{P}_{cont}$ &  8.13 $\pm$ 10.07 & 3.78 $\pm$ 9.19 & 3.17 $\pm$ 6.69 &4.19 $\pm$ 7.45 & 3.77 $\pm$ 7.81 & 1.47 $\pm$ 2.97 \\
\cmidrule(lr){2-8}
\rowcolor{gtgray}
& Ground Truth &  6.33 $\pm$ 9.33 & 2.01 $\pm$ 5.46 & 1.78 $\pm$ 4.99 &3.39 $\pm$ 7.07 & 2.84 $\pm$ 7.14 & 1.31 $\pm$ 3.58 \\
\midrule
\multirow{10}{*}{TTSV}
& DiffSinger~\cite{liu2022diffsinger} & 17.79 $\pm$ 14.16 & 16.43 $\pm$ 13.99 & 17.04 $\pm$ 14.21 & -- & -- & -- \\
& StyleSinger~\cite{zhang2024stylesinger} & 11.31 $\pm$ 7.68 & 10.43 $\pm$ 8.87 & 10.05 $\pm$ 7.54 & -- & -- & -- \\
& TCSinger~\cite{zhang2024tcsinger} & 9.88 $\pm$ 7.23 & 7.61 $\pm$ 6.15 & 7.40 $\pm$ 6.26 &29.03 $\pm$ 19.31 & 31.45 $\pm$ 21.85 & 16.11 $\pm$ 17.33 \\
& Vevo1.5~\cite{vevo}  & 8.85 $\pm$ 11.09 & 8.40 $\pm$ 10.92 & 9.34 $\pm$ 9.43 &23.62 $\pm$ 14.67 & 24.91 $\pm$ 15.17 & 11.23 $\pm$ 8.04 \\
& Vevo2~\cite{zhang2026bm} & 4.66 $\pm$ 6.04 & 2.89 $\pm$ 4.96 & 4.21 $\pm$ 5.00 & 11.37 $\pm$ 9.80 & 10.84 $\pm$ 9.61 & 4.74 $\pm$ 4.48 \\
\cmidrule(lr){2-8}
& \multicolumn{7}{l}{CookVoice} \\
& \quad $S_{\mathcal{T}} + \mathcal{P}_{disc}$ &  11.71 $\pm$ 10.44 & 11.08 $\pm$ 11.79 & 10.80 $\pm$ 8.96 &19.56 $\pm$ 14.47 & 20.81 $\pm$ 15.71 & 8.79 $\pm$ 6.72 \\
& \quad $S_{\mathcal{T}} + \mathcal{P}_{cont}$ &  9.79 $\pm$ 9.26 & 9.82 $\pm$ 12.25 & 8.92 $\pm$ 7.27 &21.10 $\pm$ 17.81 & 22.17 $\pm$ 19.71 & 10.05 $\pm$ 8.91 \\
& \quad $S_{\mathcal{V}} + \mathcal{P}_{disc}$ &  11.45 $\pm$ 9.14 & 11.15 $\pm$ 10.12 & 10.01 $\pm$ 7.93 &22.53 $\pm$ 16.92 & 23.32 $\pm$ 19.20 & 10.53 $\pm$ 7.88 \\
& \quad $S_{\mathcal{V}} + \mathcal{P}_{cont}$ &  10.01 $\pm$ 9.70 & 8.65 $\pm$ 11.99 & 8.78 $\pm$ 7.83 &24.37 $\pm$ 17.23 & 24.68 $\pm$ 20.09 & 10.54 $\pm$ 6.70 \\
\cmidrule(lr){2-8}
\rowcolor{gtgray}
& Ground Truth &  5.01 $\pm$ 4.76 & 3.06 $\pm$ 5.33 & 4.99 $\pm$ 5.98 &14.22 $\pm$ 12.93 & 13.16 $\pm$ 10.92 & 7.26 $\pm$ 8.00 \\
\bottomrule
\end{tabular}
\end{table*}

\end{document}